\documentclass{aa}
\usepackage{graphicx}
\usepackage{natbib} 
\usepackage{txfonts}
\usepackage{xcolor}
\usepackage[T1]{fontenc}
\usepackage[utf8]{inputenc}
\usepackage[toc,page]{appendix}
\usepackage[colorlinks=true,linkcolor=blue,citecolor=blue]{hyperref}
\bibpunct{(}{)}{;}{a}{}{,}

\begin{document}
\nolinenumbers

\newcommand{\msun}{\mbox{$M_{\odot}$}}
\newcommand{\mstar}{\mbox{$M_c$}}
\newcommand{\fifth}{\mbox{$5^{\rm th}$}}
\newcommand{\fei}{\mbox{Fe\,{\sc i}}} 
\newcommand{\feii}{\mbox{Fe\,{\sc ii}}} 
\newcommand{\feiii}{\mbox{Fe\,{\sc iii}}} 
\newcommand{\znii}{\mbox{Zn\,{\sc ii}}} 
\newcommand{\crii}{\mbox{Cr\,{\sc ii}}} 
\newcommand{\tiii}{\mbox{Ti\,{\sc ii}}} 
\newcommand{\cii}{\mbox{C\,{\sc ii}}} 
\newcommand{\civ}{\mbox{C\,{\sc iv}}} 
\newcommand{\sii}{\mbox{S\,{\sc ii}}} 
\newcommand{\si}{\mbox{Si\,{\sc i}}} 
\newcommand{\siii}{\mbox{Si\,{\sc ii}}} 
\newcommand{\siiii}{\mbox{Si\,{\sc iii}}} 
\newcommand{\siiv}{\mbox{Si\,{\sc iv}}} 
\newcommand{\oi}{\mbox{O\,{\sc i}}} 
\newcommand{\oiii}{\mbox{O\,{\sc iii}}} 
\newcommand{\pI}{\mbox{P\,{\sc i}}} 
\newcommand{\phoi}{\mbox{P\,{\sc i}}} 
\newcommand{\mgi}{\mbox{Mg\,{\sc i}}} 
\newcommand{\mgii}{\mbox{Mg\,{\sc ii}}} 
\newcommand{\mni}{\mbox{Mn\,{\sc i}}} 
\newcommand{\mnii}{\mbox{Mn\,{\sc ii}}} 
\newcommand{\hi}{\mbox{H\,{\sc i}}} 
\newcommand{\hii}{\mbox{H\,{\sc ii}}} 
\newcommand{\hh}{\mbox{H}2} 
\newcommand{\niii}{\mbox{Ni\,{\sc ii}}} 
\newcommand{\niiis}{\mbox{Ni\,{\sc ii$^{\star}$}}} 
\newcommand{\alii}{\mbox{Al\,{\sc ii}}} 
\newcommand{\aliii}{\mbox{Al\,{\sc iii}}} 
\newcommand{\gaii}{\mbox{Ga\,{\sc ii}}}
\newcommand{\coii}{\mbox{Co\,{\sc ii}}}
\newcommand{\siiis}{\mbox{Si\,{\sc ii$^{\star}$}}} 
\newcommand{\ois}{\mbox{O\,{\sc i$^{\star}$}}} 
\newcommand{\ciis}{\mbox{C\,{\sc ii$^{\star}$}}} 
\newcommand{\feiis}{\mbox{Fe\,{\sc ii$^{\star}$}}} 

       \title{Rapid Response Mode observations of GRB 160203A: Looking for fine-structure line variability at $z=3.52$}

       \author{G. Pugliese \inst{1} 
       \and A. Saccardi \inst{2}
       \and V. D’Elia   \inst{3} \inst{,4}
       \and S. D. Vergani \inst{2,5,6}
       \and K. E. Heintz \inst{7} \inst{,8}
       \and S. Savaglio \inst{9,10,11} 
       \and L. Kaper  \inst{1} 
       \and A. de Ugarte Postigo \inst{12}
       \and D. H. Hartmann \inst{13} 
       \and A. De Cia \inst{14} 
       \and S. Vejlgaard \inst{7} \inst{,8} 
       \and J.~P.~U. Fynbo   \inst{7} \inst{,8} 
       \and L. Christensen \inst{7} \inst{,8} 
       \and S. Campana \inst{6} 
       \and D. van Rest  \inst{1} 
       \and J. Selsing   \inst{7} \inst{,8} 
       \and K. Wiersema \inst{15} 
       \and D. B. Malesani  \inst{7} \inst{,8} \inst{,16} 
       \and S. Covino  \inst{10} 
       \and D. Burgarella \inst{14}
       \and M. De Pasquale \inst{17}
       \and P. Jakobsson  \inst{8}
       \and J. Japelj \inst{1} 
       \and D. A. Kann \inst{18}\thanks{Deceased.}
       \and C. Kouveliotou \inst{19,20} 
       \and A. Rossi \inst{10}
       \and N. R. Tanvir  \inst{21}
       \and C. C. Th\"{o}ne \inst{22} 
       \and D. Xu  \inst{23}
        }

        \institute{Astronomical Institute Anton Pannekoek, University of Amsterdam, 1090 GE Amsterdam, The Netherlands  
        \and GEPI, Observatoire de Paris, Université PSL, CNRS, 5 Place Jule Janssen, 92190 Meudon, France   
        \and Space Science Data Center (SSDC) - Agenzia Spaziale Italiana (ASI), 00133 Roma, Italy  
        \and INAF - Osservatorio Astronomico di Roma, Via Frascati 33, 00040 Monte Porzio Catone, Italy   
        \and Institut d’Astrophysique de Paris, UMR 7095, CNRS-SU, 98 bis
        boulevard Arago, 75014, Paris, France  
        \and INAF - Osservatorio Astronomico di Brera, Via E. Bianchi 46, 23807  Merate (LC), Italy   
        \and  Cosmic Dawn Center (DAWN), Denmark 
        \and Niels Bohr Institute, University of Copenhagen, Jagtvej 128, 2200 Copenhagen N, Denmark  
        \and Department of physics, University of Calabria, Via P. Bucci, Arcavacata di Rende (CS), Italy  
        \and INAF - Osservatorio di Astrofisica e Scienza dello Spazio, Via Piero Gobetti 93/3, 40129 Bologna, Italy   
        \and INFN - Laboratori Nazionali di Frascati, Frascati, Italy  
        \and Artemis, Observatoire de la C\^ote d'Azur, Universit\'{e} C\^ote d'Azur, CNRS, 06304 Nice, France   
        \and Department of Physics \& Astronomy, Clemson University, Clemson, SC 29634, USA   
        \and European Southern Observatory, Karl-Schwarzschild Str. 2, 85748 Garching bei München, Germany 
        \and Physics Department, Lancaster University, Lancaster, LA1 4YB, UK  
        \and Department of Astrophysics/IMAPP, Radboud University, 6525 AJ Nijmegen, The Netherlands 
        \and University of Messina, Department of Mathematics, Informatics, Physics and Earth Sciences, Polo Papardo, Via Stagno d'Alcontres 31, 98166 Messina, Italy   
        \and Hessian Research Cluster ELEMENTS, Giersch Science Center, Max-von-Laue-Strasse 12, Goethe University Frankfurt, Campus Riedberg, 60438 Frankfurt am Main, Germany   
        \and Physics Department, George Washington University, 725 21st Street NW, Washington, DC 20052, USA   
        \and Astronomy, Physics, and Statistics Institute of Sciences (APSIS), George Washington University, Washington, DC 20052, USA 
        \and Department of Physics and Astronomy, University of Leicester, University Road, Leicester LE1 7RH, UK   
        \and Astronomical Institute, Czech Academy of Sciences, Fri\v cova 298, Ond\v rejov, Czech Republic   
        \and Key Laboratory of Space Astronomy, National Astronomical Observatories, Chinese Sciences Academy, Beijing, 100101, China. 
        } 

\offprints{G.\ Pugliese, \email{pugliese@astroduo.org}}

\date{\today}

\abstract
{Gamma-ray bursts are the most energetic known explosions. Despite fading rapidly, they give us the opportunity to measure redshift and important properties of their host-galaxies. We report the photometric and spectroscopic study of {\it Swift} GRB\,160203A at $z = 3.518$, and its host-galaxy. Fine-structure absorption lines, detected in the afterglow at different epochs, allow us to investigate variability due to the strong fading background source.}
{We obtained two optical to near-infrared spectra of the GRB afterglow with X-shooter on ESO/VLT, 18 minutes and 5.7 hours after the burst, allowing us to investigate temporal changes of fine-structure absorption lines.}
{We measured \hi\ column density $\log N({\rm \hi/cm^{-2})} = 21.75\pm 0.10$, and several heavy-element ions along the GRB sight-line in the host-galaxy, among which \siii, \alii, \aliii, \cii, \niii, \siiv, \civ, \znii \, and \feii, and {\feii}$^*$ and {\siii}$^*$ fine structure transitions from energetic levels excited by the afterglow, at the common redshift $z = 3.518$. We measure $\mbox{[M/H]}_{\rm TOT}=-0.78\pm0.13$ and a $[\rm{Zn/Fe}]_{\rm{FIT}}=0.69 \pm 0.15$, representing the total (dust corrected) metallicity and dust depletion, respectively.  
We detected additional intervening systems along the line of sight at $z=1.03$, $z=1.26$, $z=1.98$, $z=1.99$, $z=2.20$, and $z=2.83$. 
We could not measure significant variability in the strength of the fine-structure lines throughout all the observations and determined an upper limit for the GRB distance from the absorber of $d < 300$ pc, adopting the canonical UV pumping scenario. However, we note that the quality of our data is not sufficient to conclusively rule out collisions as an alternative mechanism.} 
{GRB\,160203A belongs to a growing sample of GRBs with medium resolution spectroscopy, provided by the \textit{Swift}/X-shooter legacy program, which enables detailed investigation of the interstellar medium in high-redshift GRB host-galaxies. In particular, this host galaxy shows relatively high metal enrichment and dust depletion, already in place when the universe was only 1.8 Gyr old.}
{}


\titlerunning{RRM observations of GRB 160203A}

\maketitle
\section{Introduction}
\nolinenumbers
Long Gamma-ray bursts (GRBs) (with prompt $\gamma$-ray emission duration $> 2$ s) have been predominantly shown to be associated with the final stages of the lives of massive stars \citep{Begue2015,  gehn2013, piranom2015, Cano2017}. While it is true that some GRBs with $T_{90}>2$ s have recently been shown to be the result of compact binary mergers \citep{rastinejad2022}, these are very unlikely to contaminate high redshift samples due to the typically fainter luminosities of merger-driven GRBs and longer lifetimes of their progenitors following star formation. Thanks to their high luminosities, they have been detected up to very high redshifts \citep[$z\gtrsim 8$][]{salvat2009, Tanvir2009, Cucchiara2011, Tanvir2018}, and used to measure the metal enrichment of high-\textit{z} galaxies, thereby probing the chemical-enrichment history of the universe starting from the epoch of reionisation \citep{Fynbo2006, Campana2007, Prochaska2008, Elliott2012, Tanvir2012,  Hartoog2013, Saccardi2023}. 

During the last 14 years, programs like the {\it VLT/X-shooter GRB} consortium, and more recently the {\it Stargate} collaboration \citep{selsing2019}, have provided detailed studies of the chemical composition and  molecular gas in the interstellar medium of high redshift galaxies \citep{salvat2012, thone2013, delia2014, Hartoog2015, Kruhler2015, heintz2018, Zafar2018, Bolmer2019}.

The X-shooter GTO program first, and the {\it Stargate} Consortium afterwards used the medium resolution (resolving power $\sim 8000$) VLT/X-shooter optical-NIR spectrograph and provided to date a sample of more than 120 GRB afterglows \citep{selsing2019}, including 22 GRB afterglows with metallicity measurements \citep{Bolmer2019}. They have also largely contributed to our knowledge of the chemical properties of high-z galaxies, both with sample studies \citep{Christensen2017, Wiseman2017, Heintz2019, tanvir2019}, and with detailed studies on individual GRBs \citep{Sparre2014, postigo2018, Saccardi2023}. However, the number of observed GRB host galaxies for which the metal abundance has been determined is still relatively small (about 49 GRBs), and usually limited by spectral resolution and low signal/noise ratio (S/N) \citep{Kruhler2015, Cucchiara2015}.  

The need for a larger sample has been addressed in multiple studies 
\citep{Nagamine2008, Cucchiara2016}, in which it was stressed how the long-duration GRB peculiarity of probing mainly the inner active regions of their host galaxies can be an independent way to investigate the chemical composition of high-redshift galaxies \citep{Savaglio2003, Prochaska2009, Tanvir2018, Palmerio2019}. 
In addition, relative abundances of metals with different refractory properties are indicative of conditions in the local GRB environment (see \citealt{decia2018} for an overview).  

In addition, GRB afterglow spectra have also shown that these huge photon sources can generate a UV pumping mechanism revealed by the variability of fine-structure lines \citep{Vreeswijk2007, delia2011}. The variability has been used to reconstruct the effects GRB and afterglow radiation on the absorbing regions, demonstrating that they can influence their surrounding up to typically a few hundred parsecs \citep{Ledoux2009, Delia2009, decia2012, thone2013, Vreeswijk2013, Schady2015, postigo2018}. This in turn allows to measure the distance between the GRB and the absorbing material, providing a 3D view of the medium in the GRB host galaxies.

Despite the unique insight of high-redshift star-forming regions provided by fine-structure line variation studies, the number of GRBs having suitable data for that purpose is still very limited \citep{Delia2010, Kruhler2013, delia2014, Wiseman2017, Tanvir2018, Zafar2018} due to the need to obtain time series of spectra, starting as soon as possible after the GRB trigger. Rapid Response Mode (RRM) observations were developed at VLT also driven by this specific science case. In this mode, the telescope is robotically triggered, and observations are commenced rapidly (within minutes) after a GRB, a unique feature among large-aperture telescopes. However, the unfortunate location of the VLT with respect to the South-Atlantic anomaly (which limits the number of GRB triggers from \textit{Swift} immediately observable from Paranal), coupled with the availability of the telescope/instruments, make successful RRM observations very rare. Therefore, any new observational RRM campaign of high-redshift GRBs, focusing on the variability of fine-structure lines, contributes significantly  to our knowledge of active star-forming regions in the young Universe. 
\\ 
As a part of the VLT/X-shooter GRB follow up program, here we report the spectra of the afterglow of GRB 160203A at a redshift of $z = 3.518$, at two different epochs,  18 minutes after the $\gamma$-ray detection, and 5.7 hours after the first detection. This was the first GRB detected with the Neil Gehrels \textit{Swift} Observatory (\textit{Swift} hereafter) observed by X-shooter in RRM.

In Section 2 we discuss the main features and reduction of our VLT/X-shooter data. In Section 3, we present the results of our analysis of the property of the host galaxy, the line strength analysis, the chemical abundances and metallicities, and in Section 4, we report our results on the variability of the fine-structure lines. In Section 5, we discuss the interaction between the GRB and its environment, and in Section 6, we summarise the main outcomes of our study, also comparing them with previous analyses. \\ 

We used the following cosmological parameters: ${H_0} = 71$ $\text{km s}^{-1} \text{Mpc}^{-1}$, ${\Omega_M = 0.27}$, and $\Omega_{\Lambda} = 0.73$. Time is assumed to be in the observer frame. Unless otherwise stated, throughout the paper the uncertainties are 1-$\sigma$ and the limits are 3-$\sigma$.

\section{GRB~160203A: observations}

\subsection{\textit{Swift} observations}

GRB 160203A was a long event with a duration (time in which 90\% of the counts of the prompt $\gamma$-ray emission are detected) $T_{90} = 20.2$. It was detected by the BAT (Burst Alert Telescope) instrument onboard \textit{Swift} \citep{gehrels2004} at 02:13:10 UT on the $3^\mathrm{rd}$ of February 2016 (from here on considered as $T_0$; \citealt{GCN18979}. 

The BAT light curve was characterized by a prominent central peak at $T_0 + 12$ s, preceded and followed by less intense activity, respectively, and a spectral photon index in the gamma-ray band equal to $1.93 \pm{0.20}$.
The initial observed flux recorded by the \textit{Swift} X-Ray Telescope (XRT) in the $(0.3 - 10)$ keV energy band was $1.55^{+0.23}_{-0.20} \times 10^{-12}$ erg  cm$^{-2}$ s$^{-1}$. The energy spectrum can be described by a single power law with a photon index of  $1.76^{+0.23}_{-0.16}$ in the same energy band, absorbed by a rest0frame hydrogen-equivalent column density $N_{\text H} = 1.5^{+2.6}_{-1.5} \times 10^{22}$ ${\text{cm}}^{-2}$ \citep{GCN18986}. 

Assuming a broken power law temporal decay, $F(t) \propto t^{-\alpha}$ for the $(0.3 - 10)$ keV flux, the \textit{Swift} collaboration also reported an initial decay in the XRT light curve with a slope $\alpha_1 = 3.1^{+1.0}_{-0.9}$, and a first break at $246^{+773}_{-47}$ seconds after the $\gamma$-ray detection. A subsequent decay with a slope $\alpha_2 = 0.65^{+0.11}_{-1.97}$ was followed by another break in the light curve recorded at $6808^{+4410}_{-6181}$ seconds after the $\gamma$-ray detection. The third and final decay phase is characterized by a slope of  $\alpha_3 = 1.25^{+0.22}_{-0.16}$ \citep{evans2009}.  No {\it Swift}/UVOT detection was reported, with a magnitude limit $U > 19.5$ \citep{GCN18984}. 

\subsection{Optical-NIR Photometry}

The first reported ground-based observations in the optical-NIR band of GRB 160203A were carried out with GROND \citep{Grond2008} operating at the 2.2m MPG telescope at ESO La Silla observatory in Chile, about 6 minutes after the $\gamma$-ray detection, identifying an unknown source at the position $\text{RA(J2000.0)} = \text{10:47:48.35}$, $\text{Dec(J2000.0)} = -\text{24:47:19.8}$ with a preliminary magnitude in the AB system $r' = 18.0 \pm 0.1$ \citep{GCN18980}. 

\begin{figure}
\includegraphics[width=1.03\columnwidth]{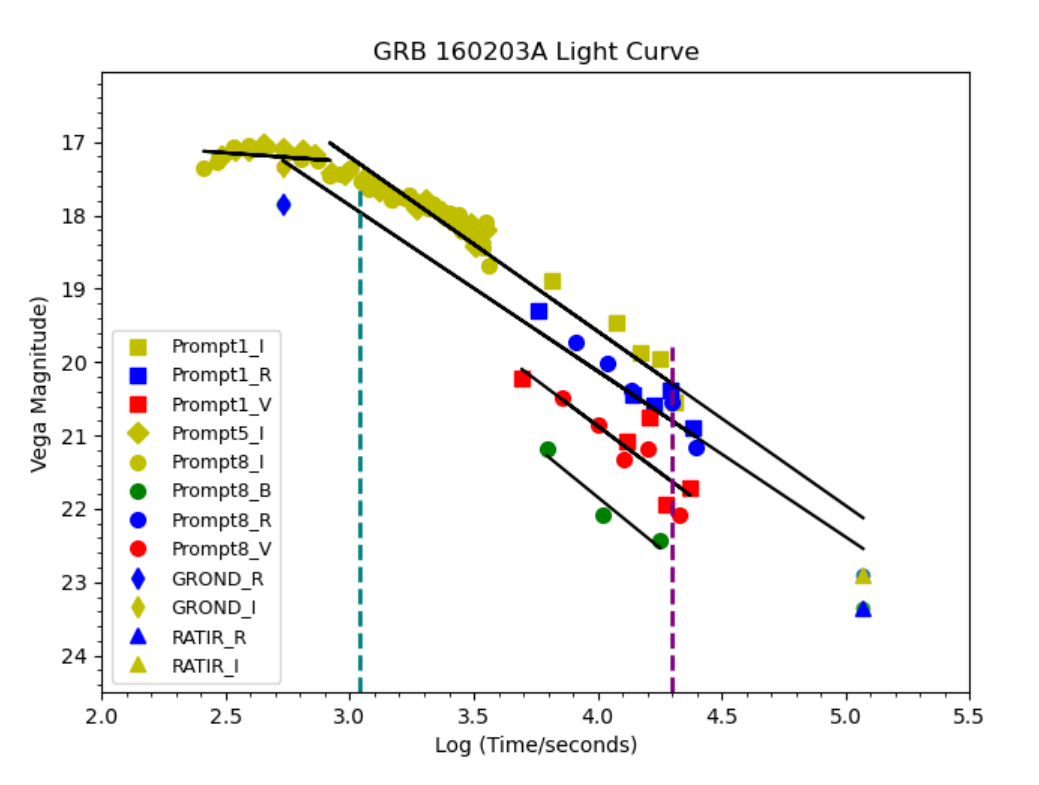}
  \caption{Optical light curve of GRB 160203A with data taken by the Skynet telescopes $16''$ telescope (all Prompt5 array data) and two $24''$ telescopes (all the Prompt1 and Prompt8 array data) at CTIO, Chile. GROND and RATIR photometric data are also included. The dashed green line corresponds to the time of our RRM observation (Epoch 1), while the purple line corresponds to the second epoch of observation (Epoch 2), both performed with X-shooter on the VLT. The black lines are the best fit for each observed band.} 
  \label{fig:grb160203a}
\end{figure}

Simultaneously with the GROND detection of the optical afterglow, the Skynet collaboration \citep{GCN18987} obtained a more extensive ground-based follow up of GRB 160203A, using their set of automated optical telescopes with apertures between 14 and 40 inches in diameter. The Skynet observations were used to create the optical light curve shown in Fig.~\ref{fig:grb160203a}. The data points were computed in the Vega magnitude systems. 
%

A later follow-up was reported by \cite{GCN18989}, using the RATIR (Reionization And Transients Infra-Red ) camera on the Harold Johnson Telescope about 32.6 hours after the $\gamma$-ray detection by \textit{Swift}. We converted their magnitude (reported in the AB system) to the Vega system (Fig.\ref{fig:grb160203a}). 
%
\begin{figure*}[ht!]
\sidecaption
   \includegraphics[width=0.69 \linewidth]{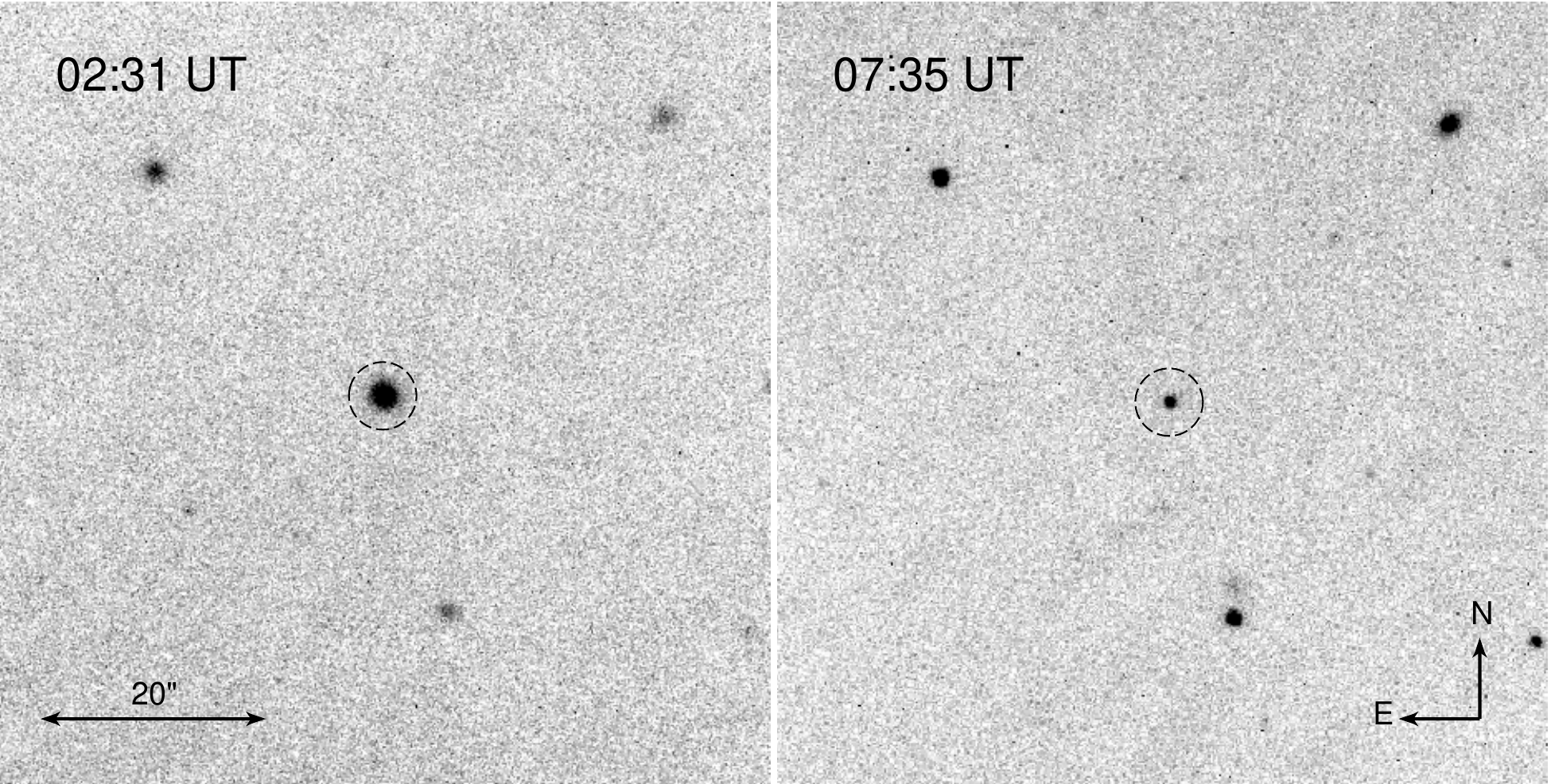}
  \caption{Acquisition images obtained in the $r$ filter during the X-shooter observations on 2016 Feb 3. 
  The left and right panels show the afterglow (circled) at the beginning of epochs 1 and 2, respectively. The fading of the GRB counterpart is apparent.}
  \label{fig:ACQ}
\end{figure*}

We performed photometry of the acquisition images obtained with X-shooter as part of the target acquisition procedure immediately prior to the spectroscopic observations (see Fig.~\ref{fig:ACQ}). These images were obtained in the $R$ filter, and consisted of two frames of 15 seconds and 5 seconds exposure time, respectively. The afterglow was observed to fade by $2.38 \pm 0.07$ magnitudes between the first and second epoch, as compared to five field stars, in agreement with the photometry collected in Fig.~\ref{fig:grb160203a}. We also performed fits to all magnitudes reported publicly in GCN circulars (Fig.~\ref{fig:grb160203a}), using a smoothly broken power law, and got indices $0.03^{+0.13}_{-0.09} $ and $1.30 ^{+1.36}_{-1.24}$ at 90\% confidence, which agrees with the X-shooter acquisition data value. There was no evidence of a jet break in the light curve of the afterglow of GRB 160203A. 

\subsection{Spectroscopic observations and data reduction}

Optical and near-infrared spectra of GRB 160203A were obtained in RRM with X-shooter \citep[first reported in][]{GCN18982} starting on $3^\mathrm{rd}$ February 02:31:35 UT, just 18 minutes after the $\gamma$-ray alert (Table \ref{Data_obs}). This was not the first time that our collaboration obtained a GRB spectrum using RRM, but this was the first time in which we succeeded to perform a second set of observations, about 5.7 hours after the alert, again using the X-shooter instrument (Table \ref{Data_obs}). \\
For clarity, we referred to all the observations taken in RRM as Epoch 1 and the observations taken 5.7 hours after the first BAT trigger, as Epoch 2.

The three X-shooter arms cover the UVB ($3000 - 5600$ {\AA}), visible ($5500 - 10200$ {\AA}) and NIR ($10200 - 24800$ {\AA}) bands simultaneously \citep{Vernet2011}. 
The observation was obtained with a slit width of 1.0, 0.9 and 0.9 arcsec in the UVB, VIS, and NIR arms, respectively, and a nominal resolving power $R = \lambda / \Delta \lambda$ of $R=5400$, $R=8900$, $R=5600$.  These $R$ values correspond to a velocity resolution of about 56, 34, and 54 $\text{km s}^{-1}$ in the UVB, VIS, and NIR arms, respectively. Velocity structure is typically used as a width of the gaussian function, but, there is no need to resolve a gaussian function, since the model is fixed.

The data reduction of the whole spectrum was performed as part of a consistent and global reduction of all the X-shooter data available with our program until 2018. A full description of the data reduction strategy is given in \cite{selsing2019}. More specifically, RRM data were acquired and reduced in ``STARE'' mode, while the second epoch observations were acquired and reduced in ``NOD'' mode \citep{Goldoni2006,Modigliani2010}. Wavelengths were corrected to the vacuum-heliocentric system.


Table~\ref{Data_obs} reports the log of the X-shooter observations, including the main observational parameters for both Epoch 1 (each set of data with a different exposure time), and Epoch 2 (each exposure had the same total exposure time of 2400 s). 
As shown in Table~\ref{Data_obs}, the S/N of the first RRM observation of Epoch 1 is very low, so they were not included in any of the analysis reported in the following Sections. 
%
\begin{table}[h]
\caption{Log of the X-shooter observations} 
\small
    \centering
    \begin{tabular}{cccccc} \hline 
\hline
 &  & Epoch 1 &  &   & \\ \hline \hline
Arm & Elapsed time & $\mathrm{Exptime}$ & Airmass & Seeing  & S/N \\
    &       (s)              &   (s)            &        &   & \\
    \hline \hline
UVB & RRM1 \, 1109 & 180 & {1.8} & {0.9} & {4.3} \\
VIS & RRM1 \, 1115 & 180 & {1.8} & {0.9} & {5.6} \\
NIR & RRM1 \, 1117 & 180 & {1.8} & {0.9} & {3.7} \\ \hline
UVB & RRM2 \, 1301 & 300 & {1.7} & {1.0} & {7.6} \\
VIS & RRM2 \, 1385 & 300 & {1.7} & {1.0} & {8.0} \\
NIR & RRM2 \, 1390 & 300 & {1.7} & {1.0} & {7.1} \\ \hline
UVB & RRM3 \, 1773 & 600 & {1.6} & {1.1} & {12.7} \\
VIS & RRM3 \, 1778 & 600 & {1.6} & {1.0} & {15.4} \\
NIR & RRM3 \, 1782 & 600 & {1.6} & {1.1} & {11.7} \\  \hline 
UVB & RRM4 \, 2465 & 1200 & {1.5} & {1.1} & {17.2} \\
VIS & RRM4 \, 2470 & 1200 & {1.5} & {1.1} & {19.9} \\
NIR & RRM4 \, 2473 & 1200 & {1.5} & {1.0} & {14.6} \\ \hline  
UVB & RRM5 \, 3758 & 1920 & {1.4} & {0.9} & {20.1} \\
VIS & RRM5 \, 3763 & 1920 & {1.4} & {0.9} & {22.4} \\
NIR & RRM5 \, 3766 & 1920 & {1.4} & {1.0} & {11.3} \\ \hline \hline 
 &  & Epoch 2 &  &   & \\ \hline \hline
UVB & 19750 & 2400 & {1.0} & {1.1} & {4.3} \\
VIS & 19758 & 2400 & {1.0} & {1.0} & {5.8} \\
NIR & 19761 & 2400 & {1.0} & {1.0} & {3.3} \\ \hline
    \end{tabular} \newline 
    \tablefoot{The table includes both Epoch 1 (RRM1, RRM2, RRM3, RRM4, and RRM5) and Epoch 2. The first column indicates the arm (UVB, VIS or NIR), the next two columns are the elapsed time after the BAT trigger and the total exposure, both in seconds. The last three columns show the average airmass, seeing, and S/N per pixel (0.2 {\AA} in the UVB and VIS arms, 0.6 {\AA} in the NIR arm) respectively.} 
    \label{Data_obs}
\end{table}

%
%

\section{Results} 

Abundances of heavy elements can reveal detailed information about the metal and dust content of the ISM along the GRB sight line in the host galaxy, and also probe the close environment in which GRBs occur. \\ 
A preliminary analysis of the spectrum of GRB 160203A was included in \cite{Bolmer2019}, as part of a sample study looking for molecular hydrogen in the X-shooter GRB afterglow spectra. In this work, we focused on a more detailed study of the chemical abundances, metallicity of both low- and high-ionization absorption lines, together with the line variability of fine-structure lines. 

\subsection{Spectroscopic analysis}
\label{Spe}

Fig.~\ref{fig:Spectral_lines} shows the X-shooter spectrum of GRB 160203A. The blue spectrum is the average of RRM3, RRM4 and RRM5 (combined to obtain the best S/N spectrum, hereafter called {\it Best Data}), whereas the red spectrum refers to Epoch 2 (the NIR-arm part of the spectrum has been rebinned for clarity). 
Identification of intervening systems were not included to avoid confusion. Some of the resonance and fine-structure lines are shown, labelled in black and red, respectively. The prominent damped Lyman-$\alpha$ absorption is visible on the left part of the top panel. 
In the top panel, both the UVB and initial spectrum in the optical band are shown, in the second panel there are most of the high-ionisation lines in the optical band, in the third panel the last part of the optical spectrum and the absorption lines in the NIR spectrum are shown, and in the bottom panel, most of the absorption lines in the NIR are visible. For clarity, not all absorption line identifications are reported.

We fit the data with the \texttt{Astrocook} code \citep{Cupani2020}, a \texttt{Python} software environment to model spectral features, both in emission and absorption (with continuum and complex absorption systems). \\

%
\begin{figure}[h!]
 \includegraphics[width=1. \linewidth]{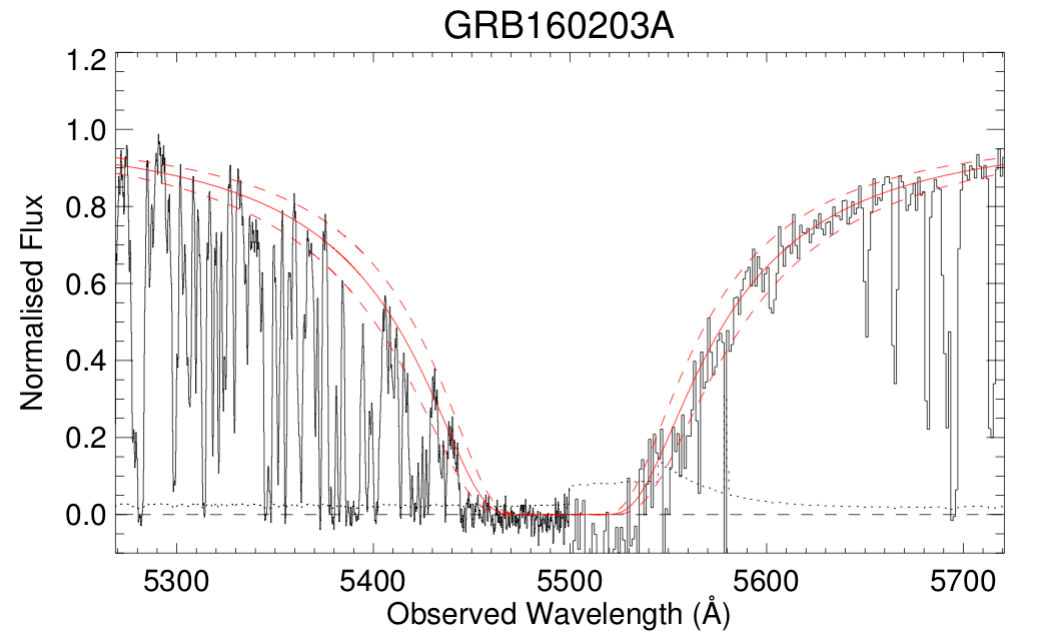}
  \caption{Best fit to the damped Ly$\alpha$ absorption line gives an \ion{H}{i} column density of $\log{\text ({N(\ion{H}{i})}} / \mathrm{cm^{-2}}) = 21.75\pm0.10$. The noise spectrum is also shown as a dotted line.}
  \label{fig:Lalpha}
\end{figure}

%
\begin{figure*}[h!]
   \includegraphics[width=0.73\textwidth]{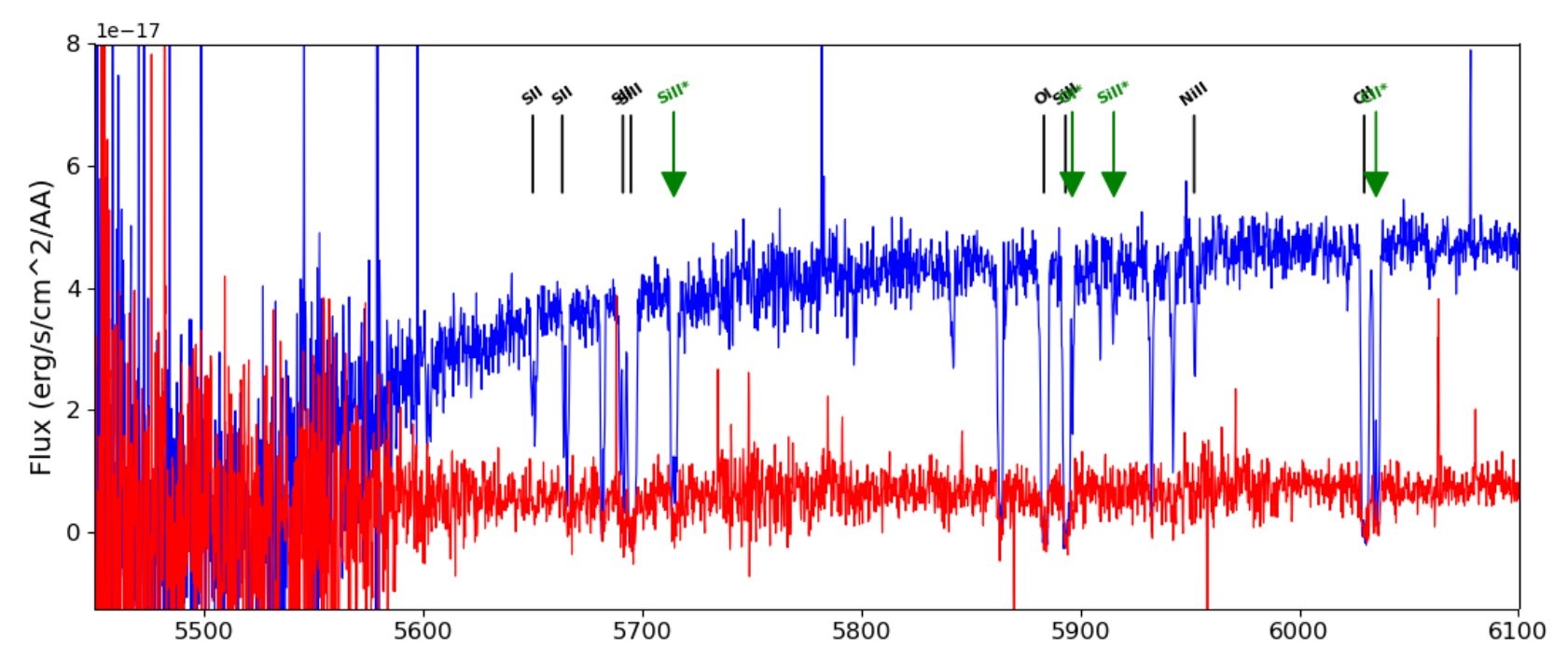}
   \includegraphics[width=0.73\textwidth]{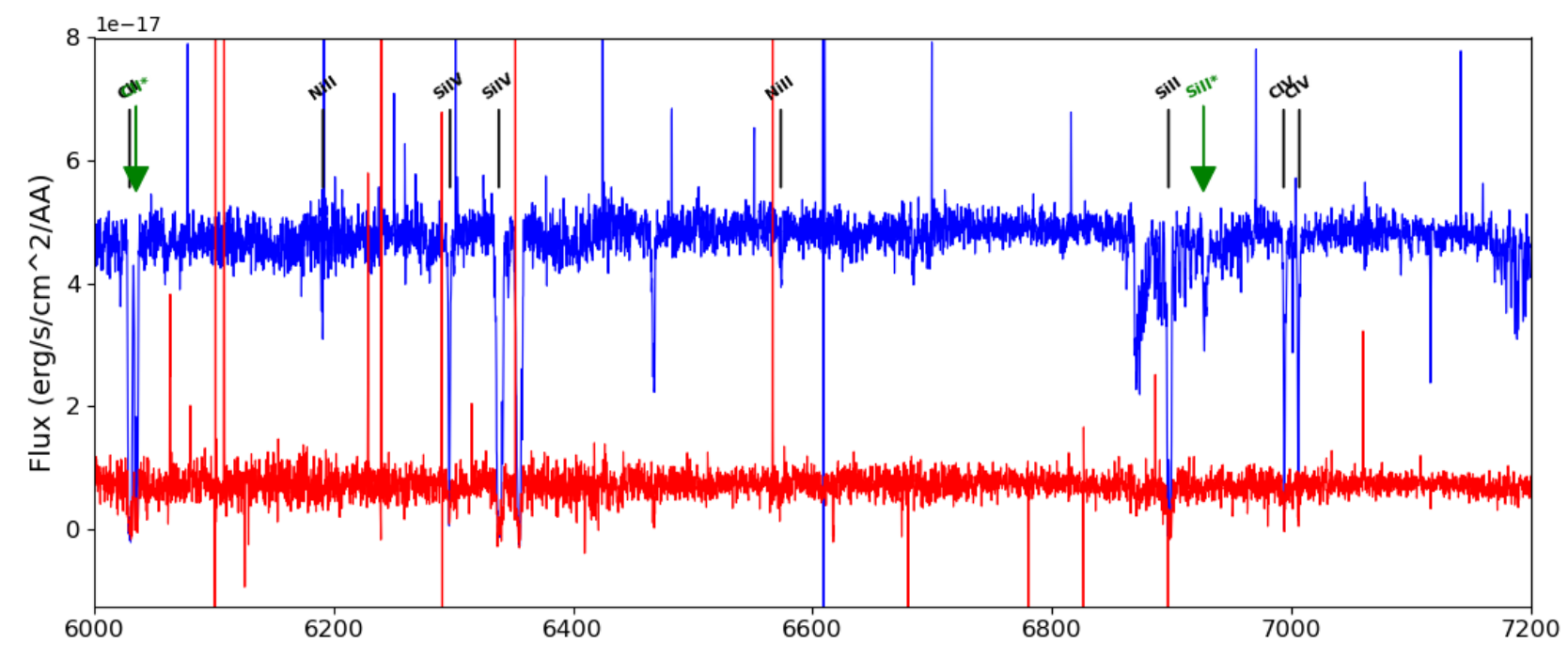} 
   \includegraphics[width=0.73\textwidth]{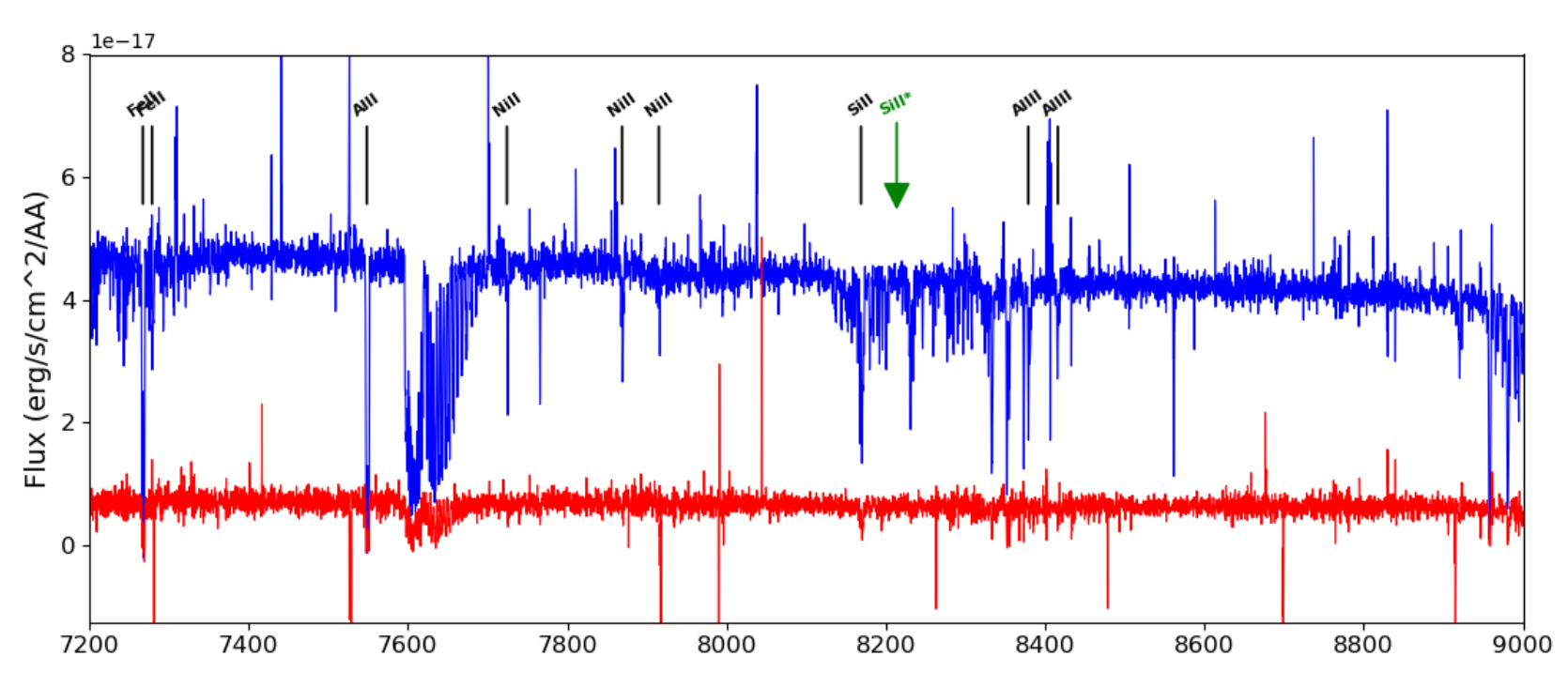}
   \includegraphics[width=0.73\textwidth]{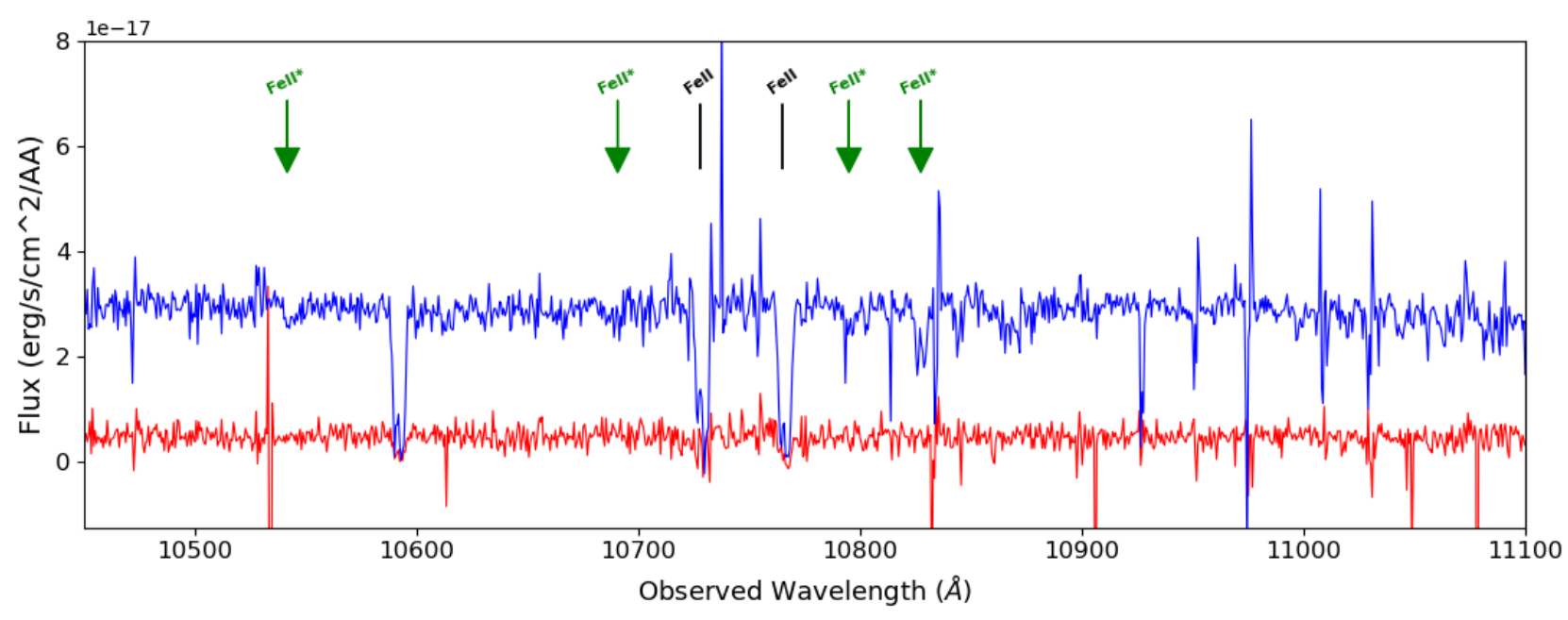}
   \centering 
  \caption{Flux desnity of the UVB, VIS and NIR X-shooter spectra for the two epochs. The average of RRM3, RRM4 and RRM5 ({\it{Best Data}}) is in blue and the Epoch 2 is in red. Line identifications are also marked, with in green the fine-structure features.}
  \label{fig:Spectral_lines}
\end{figure*}

The Ly-$\alpha$ line associated with the neutral hydrogen was located in the transition region between the UVB and VIS arms, but fortunately the HI column density was so large that both the blue and red wings of the line can be recovered. The part of the spectrum in Fig.~\ref{fig:Lalpha} shows a strong damped Ly$\alpha$ absorption (DLA, $N$(\ion{H}{i}) $> 2\times 10^{20}$\,cm$^{-2}$), and the red side of the Ly-$\alpha$ line constrained the best fit of the damped profile, from which we inferred an \ion{H}{i} column density of $\log{\text ({N(\ion{H}{i})}} / \mathrm{cm^{-2}}) = 21.75\pm 0.10$. We could not report any evidence for H$_2$ absorption lines. 

We also identify in the spectrum several absorbing systems. We associated the highest redshift system at z = 3.518 ($z_{GRB}$) with the GRB\,160203A host galaxy, given the presence of a damped Lyman-$\alpha$ (DLA) absorption and of fine-structure absorption lines. More specifically, the Voigt line profile computed for low-ionization absorption lines of GRB 160203A showed at least two velocity components at $z=3.5176$ and $z=3.5189$ ($\Delta v = 80$\,km\,s$^{-1}$), as shown in both Fig. \ref{fig:HigIon} and Appendix \ref{high_low_ion}. 
We took the redshift of the blue component ($z=3.5176$) as that of the GRB 160203A host galaxy. High-ionization lines, like  \ion{C}{iv}, and \ion{Si}{iv} are narrower, with still two components like the low-ionization lines, but the red component is much weaker, as reported in Fig. \ref{fig:HigIon}.

At the GRB redshift, we detected low-ionization absorption lines, such as \ion{S}{ii}, \ion{Si}{ii}, \ion{C}{ii}, \ion{O}{i}, \ion{Ni}{ii}, \ion{Al}{ii}, \ion{Zn}{ii}, \ion{Cr}{ii},  \ion{Mn}{ii}, \ion{Fe}{ii}, \ion{Mg}{ii} and \ion{Mg}{i}. We also detected the presence of absorption features of highly ionized species, such as \ion{Al}{iii}, \ion{C}{iv} and \ion{Si}{iv}, and fine structure lines such as \siiis, \ois, \ciis, \feiis \, together with the reported resonance lines. A complete list of all the identified absorption lines and their corresponding equivalent width (EWs) in the rest frames are reported in Table \ref{tab:redshift}. The EW errors were estimated using the formula in \cite{Cayrel1988}.

In addition, we recognised six intervening systems along the GRB sightline at $z=1.03$, $1.26$, $1.98$, $1.99$, $2.20$, and $2.83$. Due to the presence of several intervening systems, we also identified many blends with the absorption lines of the GRB host galaxy: \ion{Si}{ii}$\lambda 1260$\,\AA{} at $z_{GRB}$ is blended with \ion{Mg}{ii}$\lambda 2803$\,\AA{} at $z=1.03$, \ion{O}{i}$\lambda 1302$\,\AA{} at $z_{GRB}$ is blended with \ion{Mn}{ii}$\lambda2594$\,\AA{} at $z=1.26$, \ion{Si}{ii}$\lambda 1304$\,\AA{} at $z_{GRB}$ is blended with \ion{Fe}{ii}$\lambda 2600$\,\AA{} at $z=1.26$, \ion{Si}{iv}$\lambda 1402$\,\AA{} at $z_{GRB}$ is blended with \ion{Mg}{ii}$\lambda 2796$\,\AA{}  at $z=1.26$, \ion{Ni}{ii}$\lambda 1709$\,\AA{} at $z_{GRB}$ is blended with \ion{Fe}{ii}$\lambda 2586$\,\AA{}  at $z=1.99$ and \ion{Zn}{ii}$\lambda 2062$\,\AA{} at $z_{GRB}$ is blended with \ion{Cr}{ii}$\lambda 2062$\,\AA{} at $z_{GRB}$.

\subsection{Column densities and metallicity}
\label{columnsandmet}

While many spectroscopic analyses of long GRBs were performed between $2<z<3$ \citep{Kruhler2015, Zafar2018, Bjornsson2019, Gatkine2019}, the number of observations of long GRBs at $z>3.5$ is more limited. Detailed analysis of the two spectra of the GRB 160203A, separated by a few hours, were not only used to investigate the chemical state of the interstellar medium of the host, but also to look for fine structure lines at high redshift and their possible temporal variability.  

To compute the column densities and metallicity associated with the observed absorption lines in GRB 160203A, we used the same code, \texttt{Astrocook}, introduced in Sect. \ref{Spe}.
The total column densities of low and high ionization features, obtained from the line fitting are reported in Table \ref{Data_RRMs_ColumDens}, for the RRM2, RRM3, RRM4, RRM5  data and for Epoch 2. In Appendix \ref{CoDen_single_obs}, we reported the separated Tables (\ref{Column_Dens_RRM2}-\ref{Column_Dens_RRM5}) for each RRM observation, and in Table \ref{Column_Dens_Epoch2} the one for Epoch 2. More specifically, there are the column densities for each component (together with the total) of the low ionization absorption lines in the top panel and the high ionization absorption lines in the bottom panel for RRM2, RRM3, RRM4, RRM5 and Epoch 2 observations, respectively. 
In all tables, the Doppler parameter $b$ is provided for each component, different for the low and high ionization lines. 

%
\begin{table}[h]
\caption{List of identified absorption lines and rest-frame equivalent widths}
\footnotesize
    \centering 
    \begin{tabular}{lcc}
    \hline 
Absorption line & EW$_r$ ({\AA}) & EW$_r$ ({\AA}) \\ 
(wavelength in \AA) & \it{Best Data}  & Epoch 2 \\ 
\hline \hline 
\sii \,\,     $\lambda$1250.58 &0.24 $\pm$ 0.03     & 0.13 $\pm$ 0.08 \\
\sii \,\,     $\lambda$1253.52 &0.37 $\pm$ 0.03     & 0.26 $\pm$ 0.08 \\
\sii \,\,     $\lambda$1259.52 &0.40 $\pm$ 0.03     & 0.29 $\pm$ 0.10 \\
\siii$^a$ \,\, $\lambda$1260.42 &$<$ 0.99 & $<$ 0.80 \\
\oi$^a$ \,\,   $\lambda$1302.17 &$<$ 0.83 & $<$ 0.74 \\
\siii$^a$ \,\, $\lambda$1304.37 &$<$ 0.89 & $<$ 0.78 \\
\niii \,\,    $\lambda$1317.22 &0.09 $\pm$ 0.03     & 0.05 $\pm$ 0.08 \\ 
\cii  \,\,    $\lambda$1334.53 &1.38 $\pm$ 0.04     & 1.05 $\pm$ 0.10 \\
\niii \,\,    $\lambda$1370.13 &0.36 $\pm$ 0.04     & 0.23 $\pm$ 0.10 \\
\siiv \,\,    $\lambda$1393.76 &0.31 $\pm$ 0.04     & 0.21 $\pm$ 0.10 \\ 
\siiv$^a$ \,\, $\lambda$1402.77 &$<$ 1.26 & $<$ 1.18 \\
\niii \,\,    $\lambda$1454.84 &0.08 $\pm$ 0.04     & 0.06 $\pm$ 0.10 \\   
\siii \,\,    $\lambda$1526.71 &0.89 $\pm$ 0.04     & 1.06 $\pm$ 0.15 \\ 
\civ \,\,     $\lambda$1548.20 &0.34 $\pm$ 0.03     & 0.21 $\pm$ 0.10 \\
\civ  \,\,    $\lambda$1550.77 &0.29 $\pm$ 0.03     & 0.22 $\pm$ 0.10 \\
\feii \,\,    $\lambda$1608.45 &0.71 $\pm$ 0.03     & 0.50 $\pm$ 0.10 \\
\feii  \,\,   $\lambda$1611.20 &0.13 $\pm$ 0.04     & 0.07 $\pm$ 0.10 \\
\alii  \,\,   $\lambda$1670.79 &0.95 $\pm$ 0.03     & 0.87 $\pm$ 0.10 \\
\niii \,\,    $\lambda$1709.60 &0.12 $\pm$ 0.04     & 0.08 $\pm$ 0.10 \\
\niii  \,\,   $\lambda$1741.55 &0.16 $\pm$ 0.03     & 0.17 $\pm$ 0.10 \\ 
\niii  \,\,   $\lambda$1751.92 &0.11 $\pm$ 0.03     & 0.08 $\pm$ 0.10 \\ 
\siii  \,\,   $\lambda$1808.01 &0.55 $\pm$ 0.05     & 0.42 $\pm$ 0.10 \\
\aliii \,\,   $\lambda$1854.72 &0.26 $\pm$ 0.03     & 0.11 $\pm$ 0.10 \\
\aliii \,\,   $\lambda$1862.79 &0.17 $\pm$ 0.03     & 0.12 $\pm$ 0.10 \\
\znii  \,\,   $\lambda$2026.14 &0.37 $\pm$ 0.04     & 0.28 $\pm$ 0.10 \\
\crii  \,\,   $\lambda$2056.25 &0.10 $\pm$ 0.04     & 0.12 $\pm$ 0.10 \\
\znii$^a$ \,\, $\lambda$2062.67 &$<$ 0.46 & $<$ 0.35 \\
\feii \,\,    $\lambda$2344.21 &1.18 $\pm$ 0.03     & 1.02 $\pm$ 0.15 \\ 
\feii \,\,    $\lambda$2374.46 &0.85 $\pm$ 0.03     & 0.68 $\pm$ 0.15 \\ 
\feii \,\,    $\lambda$2382.77 &1.29 $\pm$ 0.04     & 0.93 $\pm$ 0.15 \\ 
\mnii \,\,    $\lambda$2576.88 &0.16 $\pm$ 0.05     & 0.05 $\pm$ 0.15 \\ 
\feii  \,\,   $\lambda$2586.65 &1.21 $\pm$ 0.06     & 1.08 $\pm$ 0.15 \\
\feii \,\,    $\lambda$2600.46 &1.34 $\pm$ 0.05     & 1.26 $\pm$ 0.15 \\ 
\mnii  \,\,   $\lambda$2606.46 &0.08 $\pm$ 0.06     & 0.05 $\pm$ 0.15  \\
\mgii  \,\,   $\lambda$2796.35 &1.46 $\pm$ 0.06     & 1.04 $\pm$ 0.15 \\ 
\mgii  \,\,   $\lambda$2803.00 &1.45 $\pm$ 0.06     & 1.03 $\pm$ 0.15 \\ 
\mgi  \,\,    $\lambda$2852.96 &0.38 $\pm$ 0.05     & 0.30 $\pm$ 0.15 \\ \hline
\vspace{0.2cm}
\end{tabular} 
\tablefoot{Both identified absorption lines and rest-frame equivalent widths (EW$_r$, in {\AA}) were measured using the IRAF package for {\it{Best Data}} (average of RRM3, RRM4 and RRM5 observations) and Epoch 2. The equivalent width errors were also computed in IRAF. Lines marked as $^a$ are blends with intervening absorption lines, as specified in the main text.}
    \label{tab:redshift}
\end{table}

%
\begin{figure}[h!]
\centering
   \includegraphics[width=0.9 \linewidth]{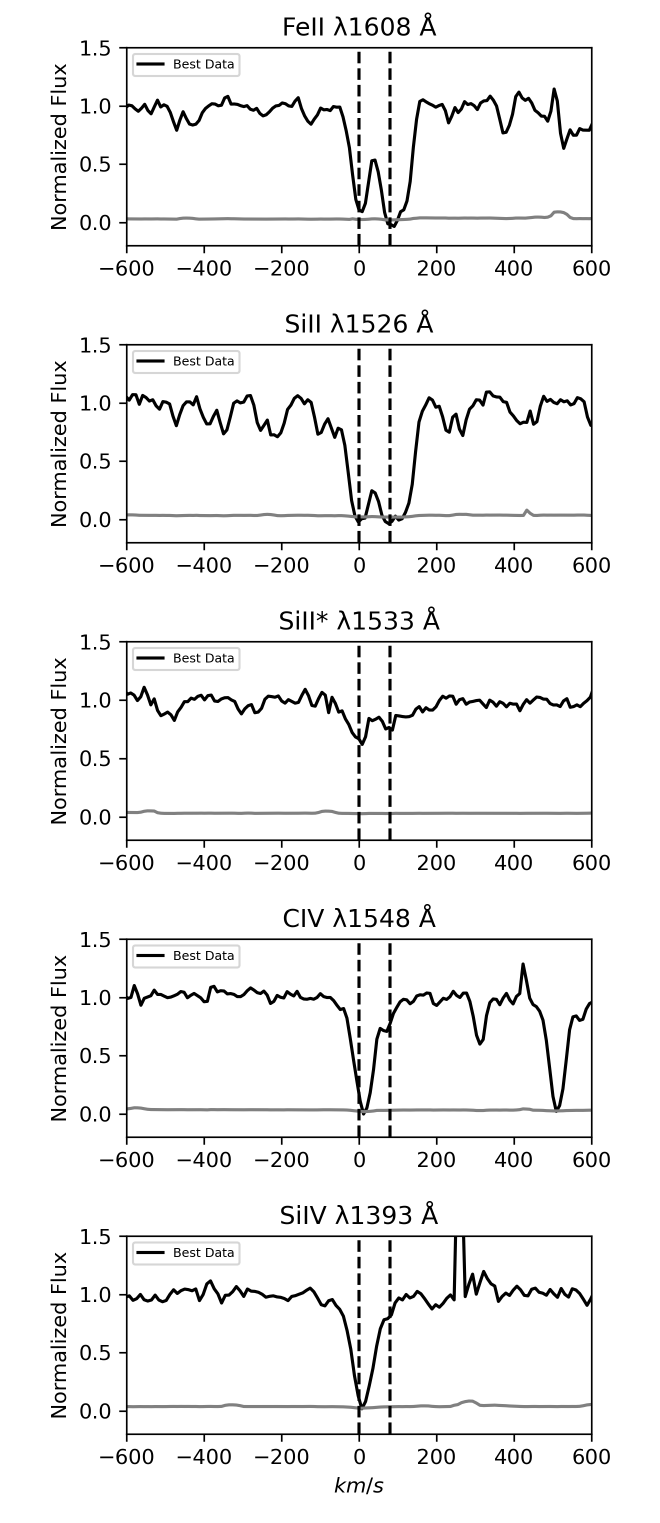}
  \caption{Selection of some low-ionization, high-ionization and fine structure lines of \ion{Fe}{II}, \ion{Si}{II}, \siiis, \ion{C}{iv} and \ion{Si}{iv}. The two vertical lines represent the redshift of the two components inferred by low-ionization lines (see Sec. \ref{Spe}). Data are in black and the error spectrum is in grey. The {\it{Best Data}} represents the average of RRM3, RRM4 and RRM5 spectra as described in Section \ref{Spe}. See Fig. \ref{Metal4} in the Appendix for all the absorption line transitions.}
  \label{fig:HigIon}
\end{figure}
We combined RRM3, RRM4 and RRM5 to obtain the best S/N spectrum ({\it Best Data}, as explained in Section \ref{Spe}), as shown in Fig.~\ref{fig:HigIon}, and a full illustration of all lines in the Appendix Fig.~\ref{Metal4}. We measured the column densities in order to calculate the relative abundances and the metallicities of the host galaxy. 
In Table \ref{Data_ColumDens}, the three columns, after the ion identification, indicate the column densities for the two strongest absorption components at velocities v\,$=0$ km~s$^{-1}$ and v\,$=80$ km~s$^{-1}$, and the total column densities, respectively. \\

%
\begin{table*}
\caption{Column densities of low and high ionization absorption lines}
\tiny
    \centering
    \begin{tabular}{lccccc}
    \hline 
Absorption lines  & $\log(N_\text{RRM2}/{\rm cm}{^{-2}})$ & $\log(N_\text{RRM3}/{\rm cm}{^{-2}})$ & $\log(N_\text{RRM4}/{\rm cm}{^{-2}})$ & $\log(N_\text{RRM5}/{\rm cm}{^{-2}})$  & $\log(N_\text{Epoch 2}/{\rm cm}{^{-2}})$ \\
\hline
\ion{Al}{ii} $\lambda$1670  & >14.0& >14.0    & >14.0    & >14.0     & >13.9\\
\ion{C}{ii} $\lambda$1334   & >15.8 & >15.9   & >16.2    & >16.1     & >15.9\\   
\ion{C}{ii$^*$} $\lambda$1335  & >15.0 & >15.0   & >15.0    & >15.1     & >15.0\\
\ion{Fe}{ii} $\lambda$1608, $\lambda1611$, $\lambda2344$, $\lambda2374$, $\lambda2382$, $\lambda2586$, $\lambda2600$  & >14.9 & >15.1    & >15.1    & >15.1     & >15.2\\ 
\ion{Fe}{ii$^*$} $\lambda$2333, $\lambda$2365, $\lambda$2389, $\lambda$2396 & 13.68  $\pm$ 0.15 & 13.64 $\pm$ 0.08    & 13.56 $\pm$ 0.07    & 13.49 $\pm$ 0.11     & 13.5 $\pm$ 0.3\\ 
\ion{Mg}{i} $\lambda$2852  & & 12.60 $\pm$ 0.09    & 12.59 $\pm$ 0.08    & 12.46 $\pm$ 0.13     &\\ 
\ion{Mg}{ii} $\lambda$2796, $\lambda$2803 & >16.0 & >15.8    & >15.9    & >16.0     & >16.0\\   
\ion{Ni}{ii}  $\lambda$1317, $\lambda$1370, $\lambda$1454, $\lambda$1709, $\lambda$1741   & & 14.06 $\pm$ 0.06    & 14.12 $\pm$ 0.05    & 14.06 $\pm$ 0.05 &\\ 
\ion{O}{i}   $\lambda$1302  & >16.0 & >16.0    & >16.2    & >16.2     & >16.1\\    
\ion{O}{i*}   $\lambda$1304 & & 13.60 $\pm$ 0.15    & 13.44 $\pm$ 0.15    & 13.97 $\pm$ 0.19     &\\ 
\ion{S}{ii} $\lambda$1250, $\lambda$1253, $\lambda$1259   & 15.61 $\pm$ 0.14 & 15.63 $\pm$ 0.10    & 15.63 $\pm$ 0.10    & 15.68 $\pm$ 0.10     & 15.8 $\pm$ 0.2\\ 
\ion{Si}{ii} $\lambda$1260, $\lambda$1304, $\lambda$1527, $\lambda$1808  & 15.97$\pm$0.07 & 16.02 $\pm$ 0.01    & 16.08 $\pm$ 0.01     &16.06 $\pm$ 0.01    &16.05$\pm$ 0.11 \\   
\ion{Si}{ii$^*$}  $\lambda1264$, $\lambda1265$, $\lambda1309$, $\lambda1533$, $\lambda1817$, $\lambda1818$ & 13.83$\pm$0.08 &14.06$\pm$0.06    &14.16$\pm$0.07     &14.15$\pm$0.05    &13.87$\pm$0.12\\
\ion{Zn}{ii}  $\lambda$2026  & 13.35 $\pm$ 0.08 & 13.34 $\pm$ 0.06    & 13.35 $\pm$ 0.02    & 13.36 $\pm$ 0.02     & 13.17 $\pm$ 0.10\\ 
\ion{Al}{iii} $\lambda$1854, $\lambda$1862  & 13.29 $\pm$ 0.07 & 13.32 $\pm$ 0.04    & 13.28 $\pm$ 0.02    & 13.22 $\pm$ 0.02     & 13.2 $\pm$ 0.2\\
\ion{C}{iV} $\lambda$1548, $\lambda$1550    & >14.4 & >14.8    & >14.6    & >14.6     & >14.5\\   
\ion{Si}{iV} $\lambda$1393, $\lambda$1402   & >14.1 & >14.1    & >14.4     & >14.4    & >14.4\\   
\hline
\hline
    \end{tabular}
    \tablefoot{Column densities for the RRM2, RRM3, RRM4 and RRM5 data set (second, third, fourth and fifth columns, respectively), and for the observation taken in Epoch 2 (last column).} 
       \label{Data_RRMs_ColumDens}
\end{table*}
%
The Doppler parameter for low ionization lines was derived from unsaturated transitions in the spectrum (e.g., \ion{Zn}{ii}$ \,\lambda 2026$\,\AA{}) and fixed for all other ions. Furthermore, by simultaneously fitting all the low-ionization absorption lines leaving the value of {\it b} as a free parameter, we got, within the errors, the same value found from the unsaturated lines. Considering the X-shooter resolution, hidden saturation cannot be excluded; on the other hand considering the flux residuals and the Doppler parameter expected for warm gas ($b\sim10~{\rm km~s^{-1}}$, as also usually found in GRB-DLAs observed at higher resolution; \citealt{Fox2008}), we estimate that hidden saturation should be marginal for some of the absorption lines detected in the spectrum. In those cases we report the respective element column densities as measurements in Tables \ref{Data_RRMs_ColumDens} and \ref{Data_ColumDens}, otherwise we report only as lower limits.

\begin{table*}
\caption{Absorption lines and column densities}
\small
    \centering
    \begin{tabular}{lcccc}
    \hline
Absorption lines  & $\log(N_{I}/{\rm cm}{^{-2}})$  & $\log(N_{II}/{\rm cm}{^{-2}})$ & $\log(N_{TOT}/{\rm cm}{^{-2}})$ \\
\hline
\ion{Al}{ii} $\lambda$1670 &>13.6 &>13.8 & >14.0\\
\ion{C}{ii} $\lambda$1334  &>16.0 &>15.6 & >16.2\\   
\ion{C}{ii$^*$} $\lambda$1335 &>14.8 &>14.8 & >15.1\\
\ion{Cr}{ii} $\lambda$2017, $\lambda$2056 &13.55  $\pm$ 0.07 &13.72  $\pm$ 0.05 & 13.94  $\pm$ 0.04\\
\ion{Fe}{ii} $\lambda$1608,  $\lambda$1611, $\lambda$2344, $\lambda$2374, $\lambda$2382, $\lambda$2586, $\lambda$2600 &>14.61 &15.01$\pm$ 0.03 &>15.14\\ 
\ion{Fe}{ii$^*$} $\lambda$2333, $\lambda$2365, $\lambda$2389, $\lambda$2396 &13.23  $\pm$ 0.11 &13.29  $\pm$ 0.09 & 13.56  $\pm$ 0.07\\ 
\ion{Mg}{i} $\lambda$2852 &12.25 $\pm$ 0.13 &12.25 $\pm$ 0.09 & 12.55 $\pm$ 0.08\\ 
\ion{Mg}{ii} $\lambda$2796, $\lambda$2803&>16.1 &>15.0 & >16.1\\   
\ion{Mn}{ii} $\lambda$2576, $\lambda$2594,  $\lambda$2606 &>12.8 &>13.3 & >13.4\\
\ion{Ni}{ii}  $\lambda$1317, $\lambda$1370, $\lambda$1454, $\lambda$1709, $\lambda$1741 &13.57 $\pm$ 0.05 &14.12 $\pm$ 0.05 & 14.22 $\pm$ 0.02\\ 
\ion{O}{i}   $\lambda$1302 &>15.8 &>15.9 & >16.2\\    
\ion{O}{i*}   $\lambda$1304 & &13.77 $\pm$ 0.19 & 13.77 $\pm$ 0.19\\ 
\ion{S}{ii} $\lambda$1250, $\lambda$1253, $\lambda$1259 &15.25$\pm$ 0.04 &15.44$\pm$ 0.04 & 15.66 $\pm$ 0.03\\ 
\ion{Si}{ii} $\lambda$1260, $\lambda$1304, $\lambda$1527, $\lambda$1808 &15.65$\pm$0.02 &15.78$\pm$0.02 & 16.02$\pm$0.02\\   
\ion{Si}{ii$^*$}  $\lambda1264$, $\lambda1265$, $\lambda1309$, $\lambda1533$, $\lambda1817$, $\lambda1818$ &13.96$\pm$0.04 &13.71$\pm$0.04 & 14.15$\pm$0.03\\
\ion{Zn}{ii}  $\lambda$2026 &12.74 $\pm$ 0.03 &13.22 $\pm$ 0.02 & 13.35 $\pm$ 0.02\\ 
\hline
$b\,{\rm (km~s^{-1})}$ & $20\pm2$ & $33\pm2$ & \\
\hline
\hline
\ion{Al}{iii} $\lambda$1854, $\lambda$1862 &13.10 $\pm$ 0.02 &12.80 $\pm$ 0.04 & 13.28 $\pm$ 0.02\\
\ion{C}{iV} $\lambda$1548, $\lambda$1550 &>14.7 &>13.3 & >14.8\\   
\ion{Si}{iV} $\lambda$1393, $\lambda$1402 &>14.5 &>12.7 & >14.5\\  
\hline
$b\,{\rm (km~s^{-1})}$ & $17\pm2$ & $26\pm2$ & \\
\hline
\hline
    \end{tabular}
    \tablefoot{Low and high ionization absorption lines in the first column, column densities of the two strongest components (second and third columns) and the corresponding total column densities of {\it Best Data}, obtained by combining the single exposure with the best S/N, i.e. RRM3, RRM4 and RRM5 observations.}
       \label{Data_ColumDens}
\end{table*}
The heavy-element abundances that we determined for the {\it Best Data} are reported in Table \ref{Data_Metall}. We derived zinc and sulfur abundance relative to hydrogen of $\mbox{[Zn/H]} = -0.96 \pm 0.11$ and $\mbox{[S/H]} = -1.21 \pm 0.10$, respectively.
We derived also zinc abundance relative to iron of $[\text{Zn/Fe}]<1.15$.
Based on the absorption line from \niii \, $\lambda 1741$ {\AA}, we computed $\Delta V_{90} = 198$ km~s$^{-1}$, which is a measure of the width of the line that contains 90\% of the optical depth \citep{Ledoux2006}. Analyses of GRB absorption lines have demonstrated that $\Delta V_{90}$ correlates with the metallicity of the host galaxy \citep{Arabsalmani2015}.
The [Zn/H] and [S/H] values are consistent with the metallicity vs.\ the $\Delta V_{90}$ relation derived by \cite{Arabsalmani2015}. 
%
\begin{figure}[h!]
\includegraphics[width=1 \linewidth]{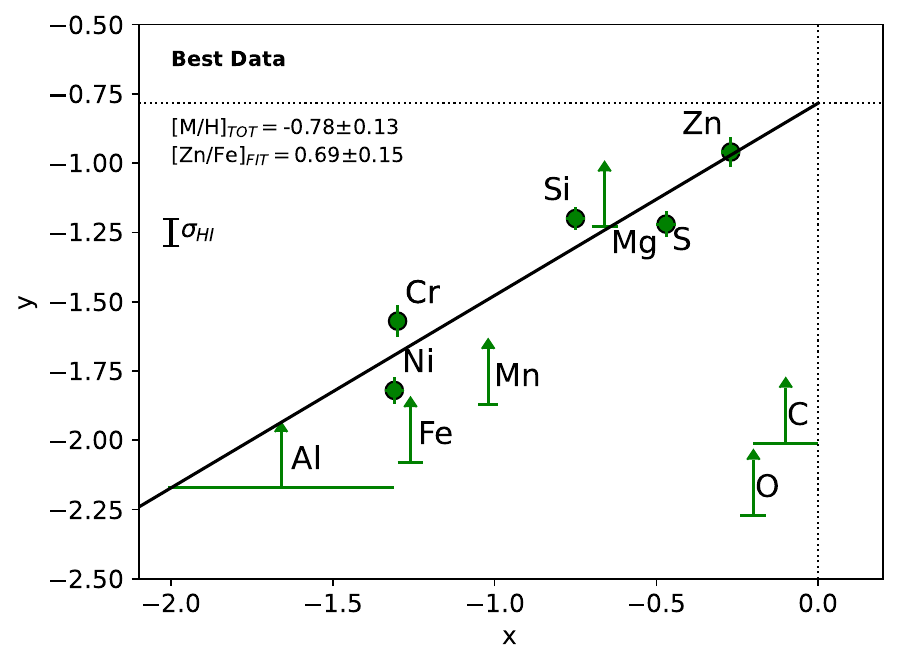}
    \caption{{\it Best data} observed abundance pattern. The $x$-axis shows the refractory index $B2_{X}$ from \citet{decia2021}, while the $y$-axis is closely related to the observed abundances. The solid line shows the linear fit of the relation $y= [{\rm Zn/Fe}]_{\rm FIT}\times x + [M/{\rm H}]_{\rm TOT}$ to the observed abundances, not including constraints from the limits. }  
  \label{Metal_Fe}
\end{figure}
%

%
\begin{table}[h!]
\caption{Heavy element abundances}
    \centering
    \begin{tabular}{lc}
    \hline 
Element   & [X/H] \\
\hline
Al   & $>-2.17$\\
C    & $>-1.98$\\
Cr   & $-1.45$ $\pm$ 0.12\\
Fe   & $>-2.09$\\
Mg   & $>-1.25$\\
Mn   & $>-1.80$\\
Ni   & $-1.75 \pm 0.11$ \\
O    & $>-2.27$ \\
S$^\dagger$  & $-1.21 \pm 0.10$\\  
Si$^\dagger$   & $-1.24 \pm 0.10$\\ 
Zn$^\dagger$    & $-0.96 \pm 0.11$\\ 
\hline
    \end{tabular}
    \tablefoot{Heavy element abundances relative to hydrogen [X/H] in the ISM measured from the GRB 160203A afterglow spectrum {\it Best Data}. $^\dagger$ Hidden saturation cannot be excluded, but it is unlikely; see Sect.\, \ref{columnsandmet}}
    \label{Data_Metall}
\end{table}

In general, the presence of dust may dramatically affect the observed abundances, because of dust depletion \citep[e.g.][]{decia2016}, so it is important to study the abundance pattern to be able to determine the total (gas+dust) metallicity, also in the neutral ISM surrounding the GRB explosion \citep[e.g.][]{Hartoog2015}. \\ 
The abundance pattern of GRB 160203A is shown in Fig. ~\ref{Metal_Fe}. 
The $x$-axis shows the refractory index $B2_{X}$ from \citet{decia2021}, which indicates the tendency of metals to be incorporated into dust grains: On the left are refractory metals, on the right the volatile ones, closer to the true metallicity. The $y$-axis is closely related to the abundances of different metals, as defined and tabulated in \cite{decia2021}, except for carbon and aluminium, which are measured in \cite{Konstantopoulou2022}.
The solid lines marks the linear fit of the relation $y= [\rm{Zn/Fe}]_{\rm FIT}\times x + [M/{\rm H}]_{\rm TOT}$ to the observed abundances, as defined by \citet{decia2021}, its normalization determines the dust-corrected total metallicity, $[M/{\rm H}]_{\rm TOT}=-0.78 \pm 0.13$, and its slope determines the overall amount of dust depletion, $[\rm{Zn/Fe}]_{\rm{FIT}}=0.69 \pm 0.15$.
We fit a linear relation to the observed abundances, considering errors along both axes and not including constraints from the limits.

\section{Fine structure line variability}

The GRB afterglow deposits a huge amount of UV radiation in the interstellar medium. This obviously impacts the physical condition of the gas along the GRB sight-line up to a certain distance, as shown by the detection of fine-structure lines in the GRB afterglow and their time variability  \citep{Prochaska2006, Vreeswijk2007, Delia2009}. The mechanism is known as indirect UV pumping and consists in UV photons exciting the gas to higher energy states. The lifetime of these states is short, therefore the atoms quickly decay to lower energy levels. The longer lived states are either excited levels with principal quantum number $n>1$, or $n=1$ states with higher values of spin-orbit coupling (the so-called fine structure levels), or a combination of the two. In particular conditions, the same mechanism could also be responsible not only for excitation, but even ionization (\citealt{Vreeswijk2013}). 

The comparison of observations with predictions from time dependent photo-excitation codes has been applied to the spectrum of several GRBs \citep{Dessau2006, Delia2009, Ledoux2009, decia2012, Hartoog2013, delia2014}. Estimated distances are between a few tens of parsec up to the kpc scale, showing that the influence of the GRB reaches remarkable distances. 

%
\begin{figure}[h!]
  \includegraphics[width=1 \linewidth]{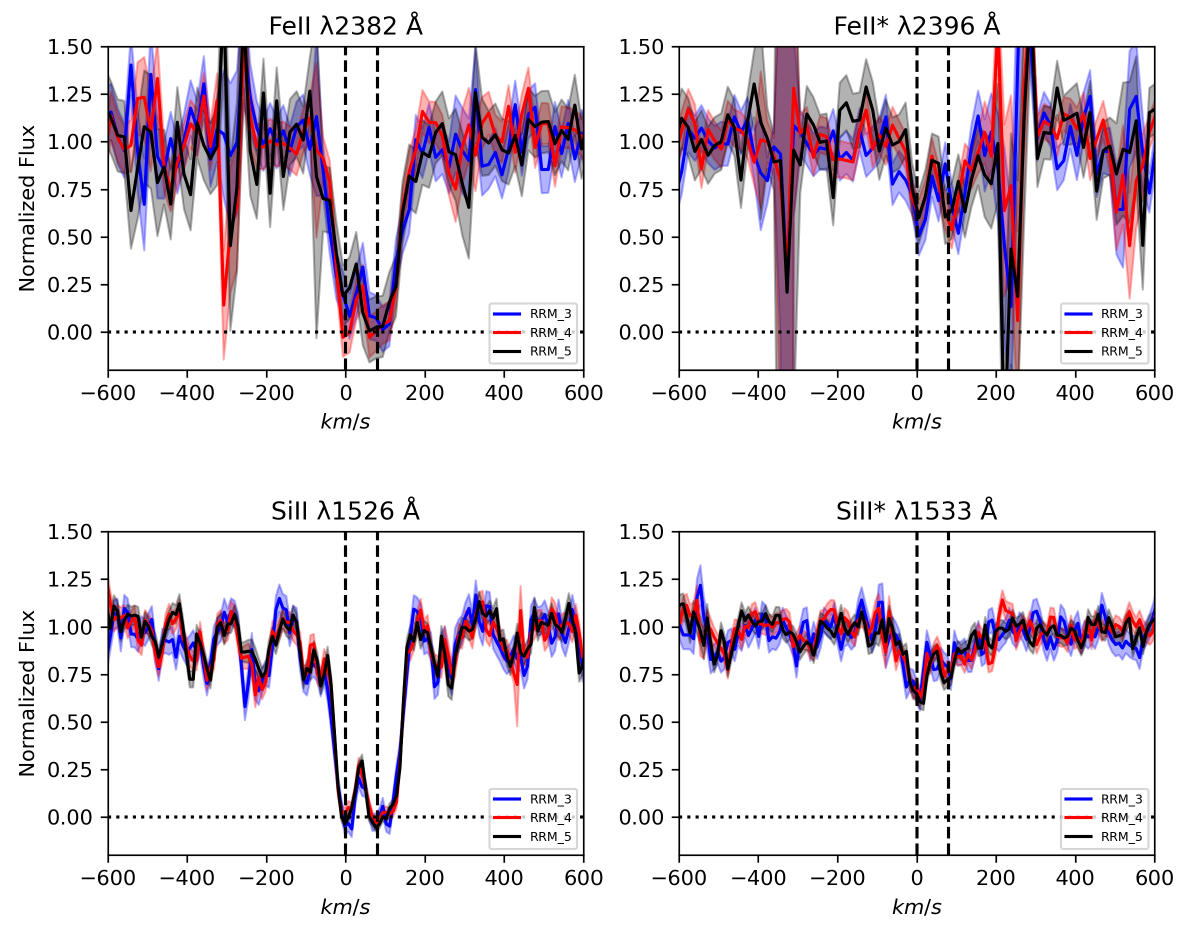}
  \caption{Resonance and fine structure lines of \ion{Fe}{ii} (top) and \ion{Si}{ii} (bottom). The blue spectrum refers to RRM3, the red spectrum refers to the RRM4 and the black line refers to the RRM5. The shaded area around the solid lines represents the error spectra for each RRM.} 
  \label{fig:fines} 
\end{figure}

{In our analysis of GRB 160203A, we noted that all the ground levels of the species commonly exhibiting fine structure lines in GRBs (\ion{C}{ii},  \ion{O}{i}, and \ion{Fe}{ii}) are saturated, apart for the \ion{Si}{ii} (see Table~\ref{Data_RRMs_ColumDens} and Table~\ref{Data_ColumDens}). For \ion{Si}{ii}, we have reliable estimate of column density for both the ground- and the fine structure features, at both epochs and both components (see Appendices). Since the fine structure levels of \ion{Si}{ii} and \ion{Fe}{ii} are compatible with being constant, as reported in Table 3, Table~\ref{Data_ColumDens}, Fig.~\ref{fig:fines} and in the Appendices, the photo-excitation code can only provide an upper limit to the source distance (see the next paragraph for a more detailed explanation). However, since the excited levels of \ion{Si}{ii} tend to remain populated even with a low flux level (see, e.g., \citealt{Saccardi2023}), it is clear that, in this case, a distance upper limit is less constraining to that obtained with \ion{Fe}{ii}. This is why in the following we focus on \ion{Fe}{ii}. 

Even if \ion{Fe}{ii} ground state features are saturated, one can nevertheless attempt to fit the fine structure column densities of the different observations, leaving the initial column density of the system as a free parameter. Results (Table~\ref{Data_RRMs_ColumDens}) indicate that these columns are compatible with being constant in the time interval between 18 minutes and 5.7 hours after the burst alert, although {\it Epoch 2} is affected by a large uncertainty (see also Table \ref{Data_FineS}, in which fine structure line EWs in the two epochs do not show strong variations). In order to obtain a nearly constant fine-structure line column density in the UV pumping scenario, the absorber must be close enough to the GRB to keep the flux level sufficiently high for a long time. 

This is because the maximum value achievable by the ratio between the  \ion{Fe}{ii} fine structure and ground levels is $8/10$. This is the ratio between the $2J+1$ quantum values of the $J=7/2$ fine structure level and $J=9/2$ for the ground one. Once this maximum is reached, no matter if the flux experienced by the absorbing gas increases, this ratio would stay the same. Thus, we check if the absorber could lie at a distance from the GRB at which the experienced flux was able to keep this ratio value at its maximum, up to the second X-shooter observation. In this way we can explain the non-variability of the fine structure line, and eventually get upper limits to the GRB-absorber distance. At larger distances, the fine structure to ground state ratio would decrease during the two observations.

This statement can be quantified with the aforementioned photo-excitation code. We adopt the \ion{Fe}{ii}$*$ column densities in Table~\ref{Column_Dens_RRM2}, \ref{Column_Dens_RRM3}, \ref{Column_Dens_RRM4}, \ref{Column_Dens_RRM5}, and \ref{Column_Dens_Epoch2}, with a range of Doppler parameters determined from absorption-line best fits and let the \ion{Fe}{ii} ground state free to vary. Physically, the latter assumption means we are assuming that the Iron contributing to the absorption is split in two regions, blended in our spectra. One of these is far from the GRB and is completely in the ground state. It contains most of the Iron. The other one is the target of the photo-excitation code. It shows the \ion{Fe}{ii}$^*$ absorption and is closer, allowing the indirect UV pumping to be at work. This region is exposed to a strong flux, allows for a \ion{Fe}{ii}$^*$/\ion{Fe}{ii} ratio close to 0.8 and consequently a (nearly) constant \ion{Fe}{ii}$*$ column density throughout the X-shooter observations. Leaving the \ion{Fe}{ii} ground column free to vary allows the model to obtain the correct normalisation to fit the data. This results in a distance upper limit, since every other fit with a lower distance is “flatter” and equally good. The above assumption was made for both components I and II of GRB 160203A. We determined upper limits to the distance between the GRB and the absorber {\bf $d < 200$ } pc and $d < 300$ pc for component I and II, respectively (see Fig.~\ref{fig:variability}).
These upper limits are at the $1\sigma$ level, and correspond to the allowed $1\sigma$ uncertainties of the measured column densities. 
These distances are consistent with values found in other GRBs (see, e.g., \citealt{Hartoog2013}), making the UV pumping a viable explanation for the presence of fine structure lines. Photoionization should not be an issue here, despite the involved distances. Indeed, GRB 080310 is the only burst for which photoionization of \ion{Fe}{ii} into \ion{Fe}{iii} was detected \citep{decia2012, Vreeswijk2013}. The peculiarity of this GRB is due to its low HI column density, with $\log{\text ({N(\ion{H}{i})}} / \mathrm{cm^{-2}}) = 18.75$. Values two orders of magnitude larger would have been high enough to allow HI to efficiently shield the Iron and prevent photoionization \citep{decia2012, Vreeswijk2013}. We note that GRB 160203A has a HI column density that is $~10^3$ times the one of GRB 080310. 

\begin{figure}[h!]
  \includegraphics[width=1.0 \linewidth]{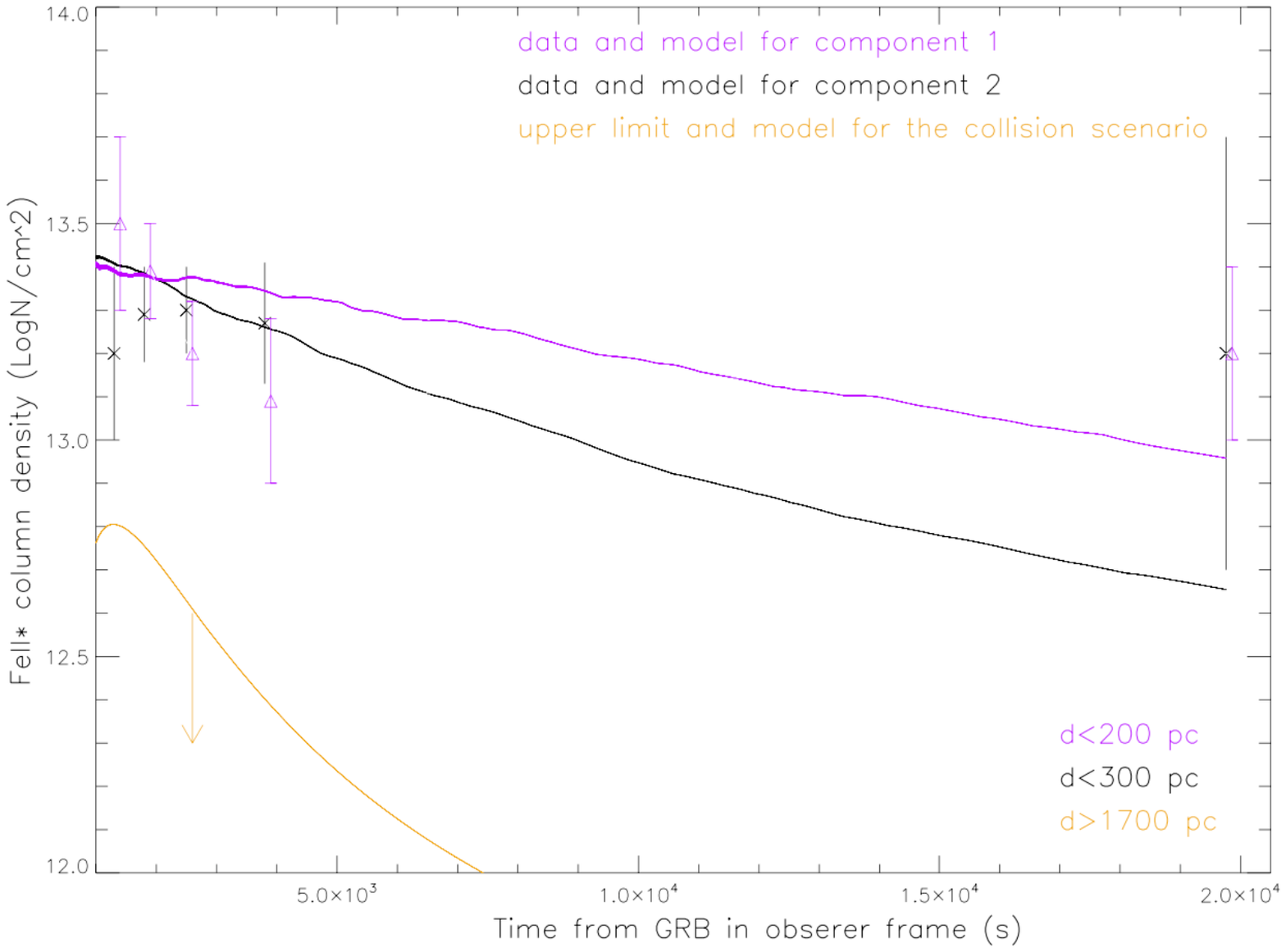}
  \caption{Comparison between the photo-excitation code and data for the \ion{Fe}{ii}$*$ column densities reported in Table~\ref{Column_Dens_RRM2}, \ref{Column_Dens_RRM3}, \ref{Column_Dens_RRM4}, \ref{Column_Dens_RRM5}, and \ref{Column_Dens_Epoch2}. We derive a GRB/absorber distance upper limit of $200$ pc and $300$ pc for Component I (purple) and II (black), respectively. For clarity reasons, purple points have been slightly shifted on the right. Assuming that fine structures are produced by collisions instead of UV pumping, we derive a lower limit to the distance of $d>1700$ pc, holding for both components (model and column upper limit are in yellow).} 
  \label{fig:variability} 
\end{figure}

We also explore the scenario in which fine structure lines are produced by collisions instead of UV pumping. Thus, we compute an upper limit to the UV pumping \ion{Fe}{ii}$*$ column density using the combined RRM3+RRM4+RRM5 spectrum, the one with the high S/N ({\it{Best Data}}). The result is Log (N \ion{Fe}{ii}$^*$/$cm^2$)$=12.6$. This value, together with the ground state column density of the  \ion{Fe}{ii} reported in Table~\ref{Data_ColumDens} for component II, was used to compare with the photo-excitation code and get a lower limit to the GRB/absorber distance in case of collisions. We obtain $d>1700 pc$ for both components (see again Fig.~\ref{fig:variability}).}

If collisions dominate over photo excitation, one can use 
the numerical solution provided by \citet{Prochaska2006} for the dependency of $n_e$ on the \ion{Fe}{ii}$^*$/\ion{Fe}{ii} ratio (together with that for the \ion{Si}{ii}$^*$/\ion{Si}{ii} one, see their fig. 9). We obtain $n_e \sim 5~$cm$^{-3}$ for \ion{Si}{ii} and $n_e \sim 200~$cm$^{-3}$ for \ion{Fe}{ii}. This discrepancy (which would hold even using a temperature different from $2600$~K, adopted by \citealp{Prochaska2006} in their fig. 9) would be even larger for a larger \ion{Fe}{ii}$^*$/\ion{Fe}{ii} ratio, which is probably underestimated (see Tables 7--10). In conclusion, the collisions scenario needs more assumptions to work, just like the photo-excitation mechanism. 

Considering that no conclusive detection of gas excited by collisions in GRBs has been reported so far, while photo-excitation has been proved to be a valid mechanism to explain the fine structure lines detected in the studies of other GRBs \textbf{\citep{Prochaska2006, Dessau2006,Vreeswijk2007, Delia2009, Ledoux2009, decia2012, Hartoog2013, delia2014}}, we prefer photo-excitation as the viable mechanism to produce the fine structure lines in GRB 160203A. Even though we need to fine tune the \ion{Fe}{ii} ground state column density, the computed distances are compatible with those reported for other bursts, and the HI column density is high enough to shield the \ion{Fe}{ii} and prevent photo-ionization into \ion{Fe}{iii}. Nevertheless, the data quality is not high enough to rule out other explanations. The collisional excitation \textbf{is} still possible, despite one needs further assumptions (as in the photo-excitation scenario), to explain the observations. 

\begin{table}[ht!]
\caption{Fine structure EW}
\small
    \centering 
    \begin{tabular}{lcc}
    \hline 
Line & EW$_r$({\AA}) & EW$_r$({\AA}) \\ 
& {\it{Best Data}}  & Epoch 2 \\ 
\hline \hline 
\siiis  $\lambda$1264.74\AA & 0.65 $\pm$ 0.03 & 0.57 $\pm$ 0.10 \\
\siiis  $\lambda$1309.28\AA & 0.12 $\pm$ 0.04 & 0.08 $\pm$ 0.11 \\ 
\ciis   $\lambda$1335.71\AA & 0.63 $\pm$ 0.03 & 0.54 $\pm$ 0.12 \\
\siiis  $\lambda$1533.43\AA & 0.22 $\pm$ 0.03 & 0.20 $\pm$ 0.12 \\
\siiis  $\lambda$1817.45\AA & 0.11 $\pm$ 0.03 & 0.09 $\pm$ 0.10 \\ 
\feiis  $\lambda$2332.02\AA & 0.31 $\pm$ 0.06 & 0.18 $\pm$ 0.12 \\ 
\feiis  $\lambda$2365.55\AA & 0.10 $\pm$ 0.07 & 0.08 $\pm$ 0.15 \\ 
\feiis  $\lambda$2389.36\AA & 0.17 $\pm$ 0.06 & 0.18 $\pm$ 0.13 \\ 
\feiis  $\lambda$2396.35\AA & 0.46 $\pm$ 0.05 & 0.28 $\pm$ 0.15 \\ 
\feiis  $\lambda$2399.98\AA & 0.19 $\pm$ 0.05 & 0.18 $\pm$ 0.15 \\ 
\feiis  $\lambda$2407.39\AA & 0.21 $\pm$ 0.06 & 0.15 $\pm$ 0.15 \\ \hline
    \end{tabular} \newline 
    \tablefoot{Rest-frame equivalent width (EW$_r$) of fine-structure lines for {\it{Best Data}} (average of RRM3, RRM4 and RRM5 observations) and Epoch 2. Errors were computed by using the formula in \cite{Cayrel1988}.}
    \label{Data_FineS}
\end{table}

\section{Discussion} 
Velocity components and measurements of line EWs can provide strong indications of the structure of the gas of the host galaxy along the line of sight to the GRB  \citep{Kupcu2007, Margutti2007, Thone2008, Arabsalmani2015, Friis2015}. The line profile computed for the ionization lines of GRB 160203A, even if saturated, shows at least two velocity components, indicating the presence of no fewer than two gas clouds along the line of sight of the afterglow. This is clearly visible in the Voigt profiles shown in the Appendix~\ref{high_low_ion} and also in Fig.~\ref{fig:fines}, in which the low ionization lines as \ion{Si}{ii} and \ion{Fe}{ii} show two distinct velocity components. The blue and the red components have a relative velocity equal to $0$ $\text{km s}^{-1}$ and $80$ $\text{km s}^{-1}$ from the $z=3.5176$ position of the blue component, respectively. 

High-ionization lines, like  \ion{C}{iv}, and \ion{Si}{iv} still show two components like the low ionization lines, with the red component much weaker, as presented in Fig.~\ref{fig:HigIon} and in the plots in Appendix~\ref{high_low_ion}. Overall, the two components of the high ionization transitions have velocity coincident with those of the low-ionization lines, and as reported in Fig.~\ref{Metal4}, the stronger component of the high-ionization lines also coincides with the stronger component of the fine structure lines, but not that of ground-state of the low-ionization lines. \\ 
Taking into consideration that both the low and high-ionization lines have consistent velocities, this could suggest that the cold and hot gas detected along the line of sight of GRB 160203A, are in a similar region of its host galaxy. 
Typically, when the high-ionization lines have a different profile than the low-ionization species, one could infer that they are associated with the circumgalactic medium (CGM), and related with turbulent motion or the coupling between gas and dust \citep{Prochaska2009, Juvela2011}. In the case of GRB 160203A, we reported high-ionization lines with a weaker component at $80$ $\text{km s}^{-1}$, and this could indicate that these species are associated with a less warm ionized gas. 

From the analysis of the column densities of the fine structure lines in both Epoch 1 and Epoch 2, we could not detect significant variability, indicating that, in the UV pumping scenario, the absorber should be close enough to GRB 160203A to allow the GRB flux being rather constant between the first and the second epoch of our observations. In detail, we computed an upper limit of $200$ pc and $300$ pc for component I and II, respectively. Such values are well within the range of what is found in the literature for similar studies. Data cannot conclusively rule out that fine structure lines are produced by collisions instead of UV pumping. In this scenario, we determined a lower limit of the GRB/absorber distance of $d>1700$ pc for both components and an electron density in the range $1-200$/cm$^3$. This wide range of n$_e$ reflects a discrepancy that arises when one uses the \ion{Fe}{ii}$^*$/\ion{Fe}{ii} or the \ion{Si}{ii}$^*$/\ion{Si}{ii} ratio. Further assumptions are needed to mitigate this discrepancy. \\ 
GRB 160203A is not the only event for which our collaboration had the opportunity to look for fine structure line variability between two following observations. GRB 100901A (\cite{Hartoog2013}) and GRB 120327A (\cite{delia2014}) are successful examples for which we could derive the distance of the absorbers from the GRB using the UV-pumping mechanism, measuring a variability within time internals of $1 - 168$ hours, and $25.62$ hours, respectively. Both these GRBs were at a lower redshift, GRB 100901A at $z=1.41$ and GRB 120327A at $z=2.81$, and for this reason they were also detected by the Ultraviolet and the Optical Telescope (UVOT) instrument on \textit{Swift} ($U=17.52$ and $U=18.02$, respectively). \\ 
Given the higher redshift, the BAT instrument on \textit{Swift} reported a lower fluence for GRB 160203A, compared to the other two events, and the \textit{Swift}/UVOT could not detect it ($U >19.5$). 
An interesting difference among these events is that both GRB 100901A and 120327A do show a variability of the fine structure features, while they are constant within the errors in GRB 160203A. This result is still consistent with an indirect UV pumping scenario, provided that the flux experienced by the intervening gas is high enough to keep these lines near to their maximum allowed value during the whole spectroscopic campaign. This requirement can be met if the intrinsic GRB luminosity is high enough and/or the absorber is sufficiently close to the GRB. 

Another goal of the spectral analysis of long GRBs was to increase the population of host galaxies at high redshift for which we were able to determine a metallicity and use this information to study the properties of the interstellar medium of late-galaxies \citep{Schady2017}. The GRB metallicities and their evolution in time are overall consistent with Quasars-DLAs \citep{Rafelski2012, decia2018}, where the decrease of metal abundance with increasing $z$ is more clear, because of the higher number of systems. The metallicity of GRB 160203A is consistent with the general indication that GRB host galaxies have lower metal abundances at higher redshift \citep{Fynbo2008, Levesque2010, Savaglio2012a, Kruhler2015}.

%

\section{Conclusion} 

As part of the VLT/X-shooter/{\it Stargate} program, we have observed several GRBs at a similar redshift as GRB 160203A. During the X-shooter science verification, GRB 090313 was detected at $z=3.373$ \citep{Postigo2010, Delia2010}, showing absorption features, \ion{S}{ii}, \ion{Si}{ii}, \ion{O}{i}, \ion{Si}{iv}, \ion{C}{iv}, \ion{Fe}{ii}, just like GRB 130408A at $z=3.758$ \citep{GCN14365} and GRB 170202A at $z=3.645$ \citep{GCN20589}. Moreover, we also observed a couple of GRBs for which we detected emission lines from the [\oiii] \, doublet, like GRB 110818A at $z=3.36$ \citep{Davanzo2011}, GRB 111123A at $z=3.151$ \citep{GCN14273}, and GRB 121201A at $z=3.385$ that showed Ly{$\alpha$} emissions \citep{GCN14035}. 
Unfortunately, we had no opportunity to perform a second set of observations within a few hours for any of these GRBs at a similar redshift, that would have allowed us to study the evolution of the detected features. 
GRB 160203A provided this opportunity, making it possible 
to check for corresponding ionization levels \citep{postigo2012}, and fine structure line variability during two close observing epochs \citep{Dessau2006, Hartoog2013}. 

From the study presented here we emphasize the following points: 
\begin{description} 
\item $-$ 
The optical and near-infrared spectra with VLT/X-shooter were obtained in RRM (5 observations, reported as Epoch 1) just 18 minutes after the $\gamma$-ray alert, together with a second set of observations (Epoch 2), about 5.7 hours after the alert. We investigated the properties of the gas along the line of sight of GRB\,160203A and we detected neutral hydrogen, low-ionization, high-ionization and fine-structure metal lines, from the GRB host galaxy at redshift $z = 3.518$. We also detected absorption lines from six intervening systems along the GRB line of sight, at $z = 1.03, 1.26, 1.98, 1.99, 2.20, 2.83.$ 
\vspace{0.06cm}
\item $-$ 
GRB 160203A shows a high \ion{H}{i} column density with respect to the DLAs of other GRBs at similar redshifts, $\log{\text ({N(\ion{H}{i})}} / \mathrm{cm^{-2}}) = 21.75\pm 0.10$, and a content of metals normal for its redshift, indicating that the region in which the GRB occurred had a high hydrogen content. The work by \cite{Ranjan2020} showed a strong similarity between the DLAs associated with long GRBs and the population of extremely strong DLAs associated with quasars with HI column densities between $\log N({\rm \hi/cm^{-2}}) = 21.6$ and $\log N({\rm \hi/cm^{-2}}) = 22.4$. The measured Hi column density of GRB 160203A, belongs to this range and gives a further indication of the similarity between the environments of the two DLA populations. 
\vspace{0.06cm}
\item $-$ 
The data show no evidence for H$_2$ absorption lines and a lack of molecular gas (see \citealt{Heintz2019} on the \ion{C}{i}-H$_2$ connection). 
\vspace{0.06cm}
\item $-$ 
We perform a detailed analysis of each observation to investigate the chemical properties of the ISM of the host galaxy and to look for fine structure line variability. We found a dust-corrected metallicity of $[\rm{M/H}]_{\rm{TOT}}=-0.78\pm0.13$, and overall strength of the dust depletion $[\rm{Zn/Fe}]_{\rm{FIT}}=0.69 \pm 0.15$.
\vspace{0.06cm}
\item $-$ 
Low-ionisation absorption lines show a width that is consistent with the GRB metallicity, while the high-ionisation lines exhibit a remarkable narrow structure compared to other GRBs previously studied, and similarly seen in the QSO-DLAs sample by \cite{heintz2018}. 
The line profile computed for both the low and high ionization lines of GRB 160203A, shows at least two components, with coincident velocities. 
\vspace{0.06cm}
\item $-$ 
From the modeling of the fine-structure lines we try to estimate the distance of the absorbing gas clouds. The small variation (if any) of the \ion{Fe}{ii} fine structure line, together with the lack of a reliable value for the corresponding ground state column density, does not allow a firm estimate of the distance at which the related absorbing gas should be located, assuming a UV pumping model. Nevertheless, an upper limit of {\bf $d<200$  }($d<300$) pc for component I (II) can be estimated. These values do not conflict with previous determinations of the same quantity for other GRBs \citep{Hartoog2013}, making UV pumping a viable explanation for the presence of fine structure lines in GRB 160203A. Nevertheless, although we prefer the photo-excitation scenario, data quality is not high enough to conclusively rule out collisions as the mechanism to produce fine structure lines. In this case, we can determine a lower limit to the GRB/absorber distance of $d>1700$ pc for both components.
\end{description} 

Overall, the observations and corresponding spectral and photometric analysis of GRB 160203A confirmed the effectiveness of using GRBs as a tool to study the chemical composition of galaxies at high redshift and highlights the value of rapid multi-epoch follow-up of GRBs.

\begin{acknowledgements}
Based on observations collected at the European Organisation for Astronomical Research in the Southern Hemisphere under ESO programme 096.A-0079, PI: J.P.U.Fynbo. This work made use of data collected by the \textit{Swift} satellite. The Cosmic Dawn Center (DAWN) is funded by the Danish National Research Foundation under grant No. 140. G.P. thanks the Swinburne University of Technology for their support as one of their seasonal teaching staff. A.S. and S.D.V. acknowledge support from CNES and DIM-ACAV+. K.E.H. acknowledges support from the Carlsberg Foundation Reintegration Fellowship Grant CF21-0103. A.D.C. acknowledges support by the Swiss National Science Foundation under grant 185692. DAK acknowledges support from Spanish National Research Project RTI2018-098104-J-I00 (GRBPhot). DBM is supported by the European Research Council (ERC) under the European Union’s Horizon 2020 research and innovation programme (grant agreement No.~725246).
\end{acknowledgements}

\newpage

\bibliographystyle{aa}
\bibliography{GRBs-refs.bib}
%
%
\onecolumn
\begin{appendices}
\appendix
\section{Voigt profiles of low and high ionization, and fine structure metal lines.}
\label{high_low_ion}

\begin{figure*}[h]
  \includegraphics[width=1 \linewidth]{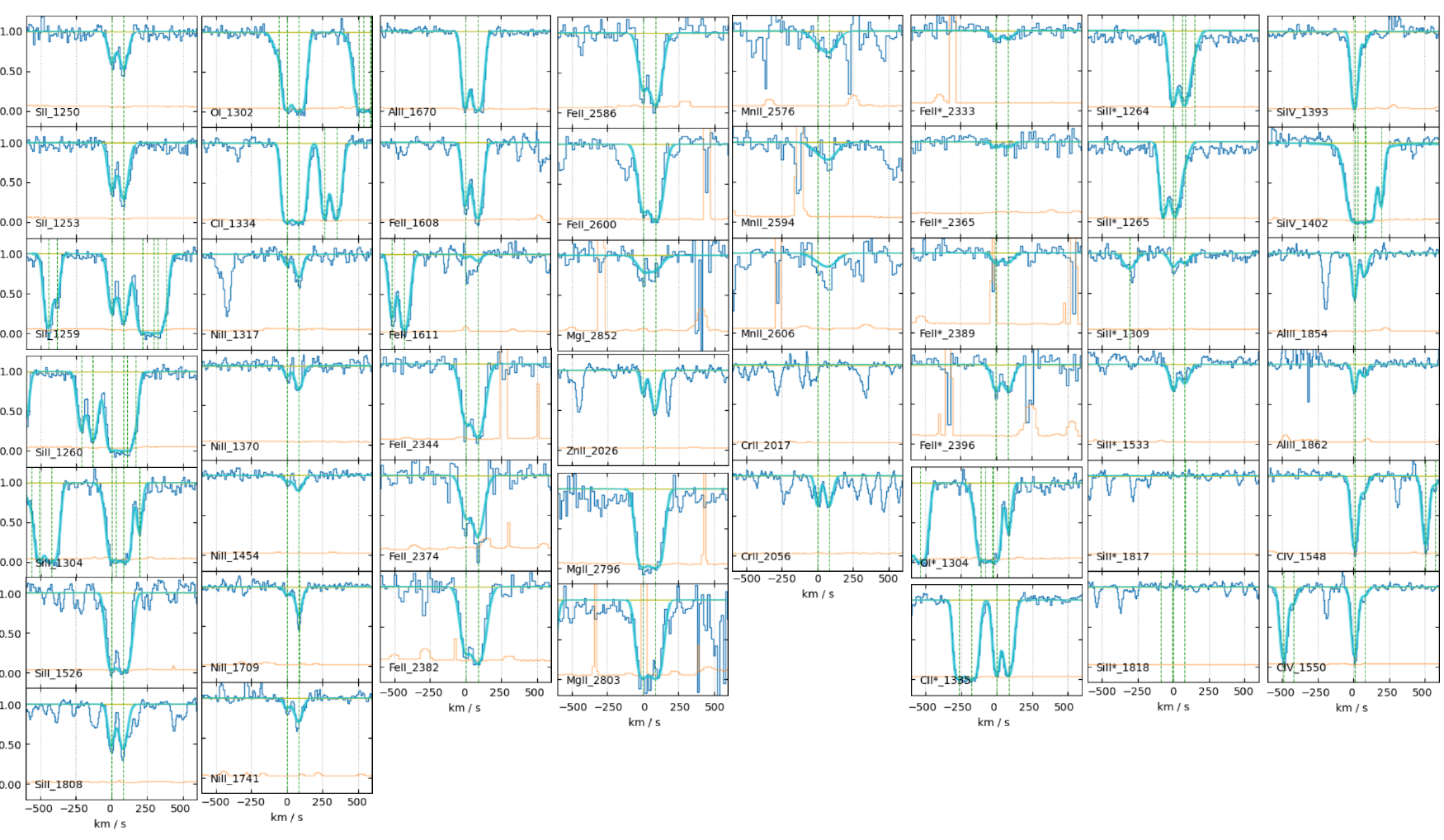}
  \caption{Set of Voigt profile fits for metal lines present in the spectrum of GRB\,160203A {\it Best Data}, by combining single exposures RRM3, RRM4 and RRM5. Data are in blue, the fit is in cyan, the error spectrum is in orange, the continuum level is in yellow, and the vertical green dashed lines indicate the center of the components.}
  \label{Metal4}
\end{figure*} 

\begin{figure*}[h]
  \includegraphics[width=1 \linewidth]{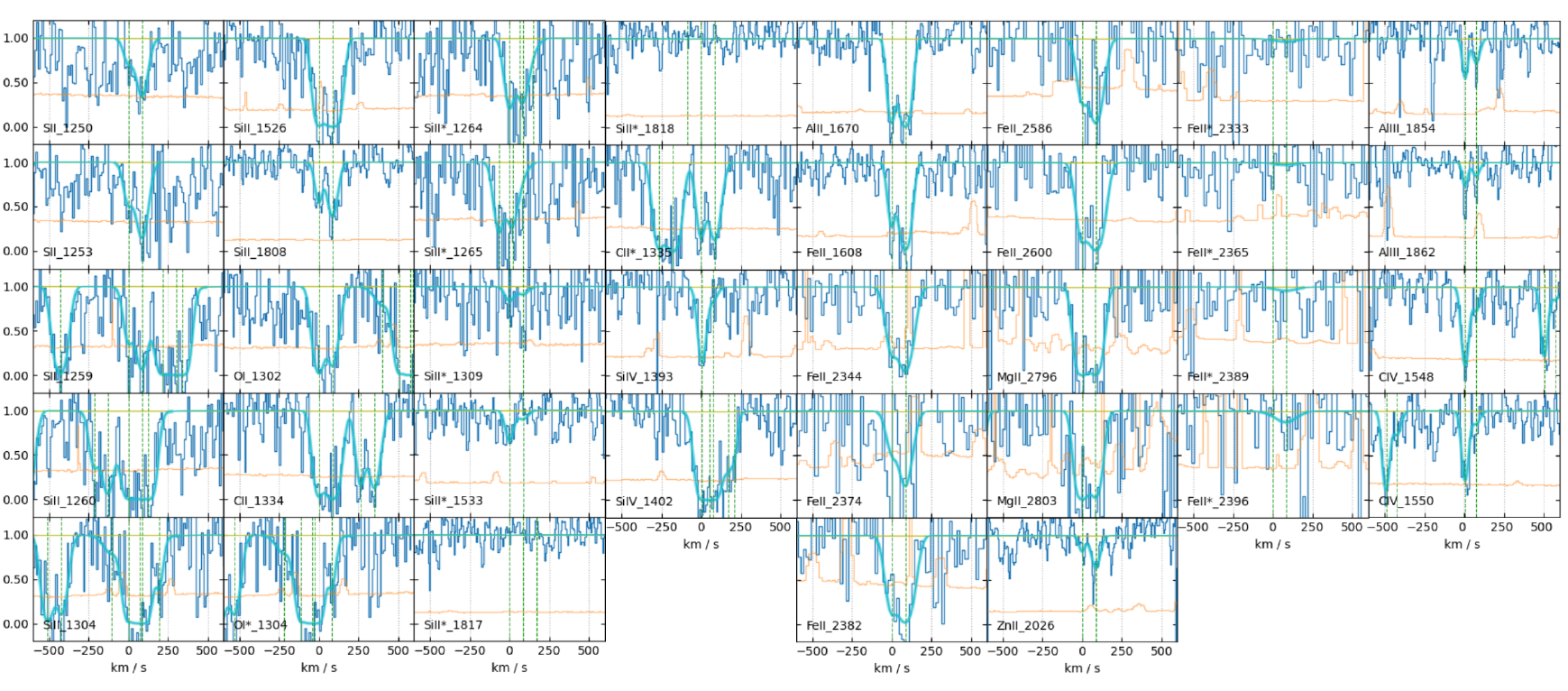}
  \caption{Set of Voigt profile fits for metal lines present in the spectrum of GRB\,160203A Epoch 2. Data are in blue, the fit is in cyan, the error spectrum is in orange, the continuum level is in yellow, and the vertical green dashed lines indicate the center of the components.}
  \label{Metal5}
\end{figure*} 

\newpage

\section{Column densities of low and high ionization absorption lines for each observation.}
\label{CoDen_single_obs}

\begin{table*}[h!]
\caption{RRM2 strongest components' column densities}
\scriptsize
    \centering
    \begin{tabular}{lcccc}
    \hline
Absorption lines & $\log(N_{I}/{\rm cm}{^{-2}})$  & $\log(N_{II}/{\rm cm}{^{-2}})$ & $\log(N_{TOT}/{\rm cm}{^{-2}})$ \\
\hline
\ion{Al}{ii} $\lambda$1670 &>13.3 &>13.9 & >14.0\\
\ion{C}{ii} $\lambda$1334  &>15.4 &>15.5 & >15.8\\   
\ion{C}{ii$^*$} $\lambda$1335 &>14.6 &>14.8 & >15.0\\
\ion{Fe}{ii} $\lambda$1608,  $\lambda$1611, $\lambda$2344, $\lambda$2374, $\lambda$2382, $\lambda$2586, $\lambda$2600 &>14.5 &14.72$\pm$ 0.15 &>14.9\\ 
\ion{Fe}{ii$^*$} $\lambda$2333, $\lambda$2365, $\lambda$2389, $\lambda$2396 &13.5  $\pm$ 0.2 &13.2  $\pm$ 0.2 & 13.68  $\pm$ 0.15\\ 
\ion{Mg}{ii} $\lambda$2796, $\lambda$2803&>16.0 &>15.3 & >16.04\\   
\ion{O}{i}   $\lambda$1302 &>15.2 &>16.0 & >16.0\\    
\ion{S}{ii} $\lambda$1250, $\lambda$1253, $\lambda$1259 &15.2$\pm$ 0.2 &15.4$\pm$ 0.2 & 15.61 $\pm$ 0.14\\ 
\ion{Si}{ii} $\lambda$1260, $\lambda$1304, $\lambda$1527, $\lambda$1808 &15.56$\pm$0.10 &15.76$\pm$0.10 & 15.97$\pm$0.07\\   
\ion{Si}{ii$^*$}  $\lambda1264$, $\lambda1265$, $\lambda1309$, $\lambda1533$, $\lambda1817$, $\lambda1818$ &13.61$\pm$0.11 &13.44$\pm$0.10 & 13.83$\pm$0.08\\
\ion{Zn}{ii}  $\lambda$2026 &12.83 $\pm$ 0.10 &13.18 $\pm$ 0.10 & 13.35 $\pm$ 0.08\\ 
\hline
\hline
$b\,{\rm (km~s^{-1})}$ & $24\pm2$ & $34\pm2$ & \\
\hline
\hline
\hline
\hline
\ion{Al}{iii} $\lambda$1854, $\lambda$1862 &13.15 $\pm$ 0.06 &12.9 $\pm$ 0.2 & 13.29 $\pm$ 0.07\\
\ion{C}{iV} $\lambda$1548, $\lambda$1550 &>14.4 &13.34$\pm$ 0.19 & >14.4\\   
\ion{Si}{iV} $\lambda$1393, $\lambda$1402 &>14.1 &13.02$\pm$ 0.11 & >14.1\\  
\hline
\hline
$b\,{\rm (km~s^{-1})}$ & $16\pm2$ & $25\pm2$ & \\
\hline
\hline
    \end{tabular}
    \tablefoot{Column densities of the two strongest components (second and third columns) and the corresponding total column densities of the low (upper panel), and high ionization (bottom panel) absorption lines of the RRM2 observation.}
       \label{Column_Dens_RRM2}
\end{table*}

\begin{table*}[h!]
\caption{RRM3 strongest components' column densities}
\scriptsize
    \centering
    \begin{tabular}{lcccc}
    \hline
Absorption lines & $\log(N_{I}/{\rm cm}{^{-2}})$  & $\log(N_{II}/{\rm cm}{^{-2}})$ & $\log(N_{TOT}/{\rm cm}{^{-2}})$ \\
\hline
\ion{Al}{ii} $\lambda$1670 &>13.4 &>13.9 & >14.0\\
\ion{C}{ii} $\lambda$1334  &>15.7 &>15.4 & >15.9\\   
\ion{C}{ii$^*$} $\lambda$1335 &>14.7 &>14.8 & >15.0\\
\ion{Fe}{ii} $\lambda$1608,  $\lambda$1611, $\lambda$2344, $\lambda$2374, $\lambda$2382, $\lambda$2586, $\lambda$2600 &>14.6 &14.98$\pm$ 0.04 &>15.1\\ 
\ion{Fe}{ii$^*$} $\lambda$2333, $\lambda$2365, $\lambda$2389, $\lambda$2396 &13.39  $\pm$ 0.11 &13.29  $\pm$ 0.11 & 13.64  $\pm$ 0.08\\ 
\ion{Mg}{i} $\lambda$2852 &12.27 $\pm$ 0.15 &12.31 $\pm$ 0.10 & 12.60 $\pm$ 0.09\\ 
\ion{Mg}{ii} $\lambda$2796, $\lambda$2803&>15.9 &>14.8 & >15.9\\   
\ion{Ni}{ii}  $\lambda$1317, $\lambda$1370, $\lambda$1454, $\lambda$1709, $\lambda$1741 &13.28 $\pm$ 0.14 &13.99 $\pm$ 0.07 & 14.06 $\pm$ 0.06\\ 
\ion{O}{i}   $\lambda$1302 &>15.6 &>15.9 & >16.1\\    
\ion{O}{i*}   $\lambda$1304 & &13.60 $\pm$ 0.15 & 13.60 $\pm$ 0.15\\ 
\ion{S}{ii} $\lambda$1250, $\lambda$1253, $\lambda$1259 &15.17$\pm$ 0.10 &15.44$\pm$ 0.06 & 15.63 $\pm$ 0.10\\ 
\ion{Si}{ii} $\lambda$1260, $\lambda$1304, $\lambda$1527, $\lambda$1808 &15.74$\pm$0.02 &15.69$\pm$0.02 & 16.02$\pm$0.01\\   
\ion{Si}{ii$^*$}  $\lambda1264$, $\lambda1265$, $\lambda1309$, $\lambda1533$, $\lambda1817$, $\lambda1818$ &13.85$\pm$0.10 &13.65$\pm$0.06 & 14.06$\pm$0.06\\
\ion{Zn}{ii}  $\lambda$2026 &12.77 $\pm$ 0.05 &13.21 $\pm$ 0.03 & 13.34 $\pm$ 0.06\\ 
\hline
\hline
$b\,{\rm (km~s^{-1})}$ & $22 \pm2$ & $32 \pm2$ & \\
\hline
\hline
\hline
\hline
\ion{Al}{iii} $\lambda$1854, $\lambda$1862 &13.09 $\pm$ 0.02 &12.94 $\pm$ 0.04 & 13.32 $\pm$ 0.04\\
\ion{C}{iV} $\lambda$1548, $\lambda$1550 &>14.7 &13.37$\pm$ 0.06 & >14.8\\   
\ion{Si}{iV} $\lambda$1393, $\lambda$1402 &>14.1 &13.02$\pm$ 0.11 & >14.1\\  
\hline
\hline
$b\,{\rm (km~s^{-1})}$ & $16\pm2$ & $25\pm2$ & \\
\hline
\hline
    \end{tabular}
    \tablefoot{Column densities of the two strongest components (second and third columns) and the corresponding total column densities of  the low (upper panel), and high ionization (bottom panel) absorption lines of the RRM3 observation.}
       \label{Column_Dens_RRM3}
\end{table*}

\begin{table*}[h!]
\caption{RRM4 strongest components' column densities}
\scriptsize
    \centering
    \begin{tabular}{lcccc}
    \hline
Absorption lines & $\log(N_{I}/{\rm cm}{^{-2}})$  & $\log(N_{II}/{\rm cm}{^{-2}})$ & $\log(N_{TOT}/{\rm cm}{^{-2}})$ \\
\hline
\ion{Al}{ii} $\lambda$1670 &>13.6 &>13.8 & >14.0\\
\ion{C}{ii} $\lambda$1334  &>16.0 &>15.9 & >16.2\\   
\ion{C}{ii$^*$} $\lambda$1335 &>14.7 &>14.6 & >15.0\\
\ion{Fe}{ii} $\lambda$1608,  $\lambda$1611, $\lambda$2344, $\lambda$2374, $\lambda$2382, $\lambda$2586, $\lambda$2600 &>14.6 &14.99$\pm$ 0.03 &>15.1\\ 
\ion{Fe}{ii$^*$} $\lambda$2333, $\lambda$2365, $\lambda$2389, $\lambda$2396 &13.20  $\pm$ 0.12 &13.30  $\pm$ 0.10 & 13.56  $\pm$ 0.07\\ 
\ion{Mg}{i} $\lambda$2852 &12.28 $\pm$ 0.12 &12.29 $\pm$ 0.09 & 12.59 $\pm$ 0.08\\ 
\ion{Mg}{ii} $\lambda$2796, $\lambda$2803&>15.9 &>14.6 & >15.9\\   
\ion{Ni}{ii}  $\lambda$1317, $\lambda$1370, $\lambda$1454, $\lambda$1709, $\lambda$1741 &13.61 $\pm$ 0.11 &13.95 $\pm$ 0.06 & 14.12 $\pm$ 0.05\\ 
\ion{O}{i}   $\lambda$1302 &>15.8 &>15.9 & >16.2\\    
\ion{O}{i*}   $\lambda$1304 & &13.44 $\pm$ 0.15 & 13.44 $\pm$ 0.15\\ 
\ion{S}{ii} $\lambda$1250, $\lambda$1253, $\lambda$1259 &15.20$\pm$ 0.10 &15.43$\pm$ 0.06 & 15.63 $\pm$ 0.10\\ 
\ion{Si}{ii} $\lambda$1260, $\lambda$1304, $\lambda$1527, $\lambda$1808 &15.70$\pm$0.02 &15.85$\pm$0.02 & 16.08$\pm$0.01\\   
\ion{Si}{ii$^*$}  $\lambda1264$, $\lambda1265$, $\lambda1309$, $\lambda1533$, $\lambda1817$, $\lambda1818$ &13.97$\pm$0.10 &13.69$\pm$0.06 & 14.16$\pm$0.07\\
\ion{Zn}{ii}  $\lambda$2026 &12.72 $\pm$ 0.04 &13.23 $\pm$ 0.02 & 13.35 $\pm$ 0.02\\ 
\hline
\hline
$b\,{\rm (km~s^{-1})}$ & $21\pm2$ & $33\pm2$ & \\
\hline
\hline
\hline
\hline
\ion{Al}{iii} $\lambda$1854, $\lambda$1862 &13.11 $\pm$ 0.02 &12.79 $\pm$ 0.04 & 13.28 $\pm$ 0.02\\
\ion{C}{iV} $\lambda$1548, $\lambda$1550 &>14.6 &13.19$\pm$ 0.07 & >14.6\\   
\ion{Si}{iV} $\lambda$1393, $\lambda$1402 &>14.4 &12.60$\pm$ 0.16 & >14.4\\  
\hline
\hline
$b\,{\rm (km~s^{-1})}$ & $17\pm2$ & $23\pm2$ & \\
\hline
\hline
    \end{tabular}
    \tablefoot{Column densities of the two strongest components (second and third columns) and the corresponding total column densities of  the low (upper panel), and high ionization (bottom panel) absorption lines of the RRM4 observation.}
       \label{Column_Dens_RRM4}
\end{table*}

\begin{table*}[h!]
\caption{RRM5 strongest components' column densities}
\scriptsize
    \centering
    \begin{tabular}{lcccc}
    \hline
Absorption lines & $\log(N_{I}/{\rm cm}{^{-2}})$  & $\log(N_{II}/{\rm cm}{^{-2}})$ & $\log(N_{TOT}/{\rm cm}{^{-2}})$ \\
\hline
\ion{Al}{ii} $\lambda$1670 &>13.6 &>13.8 & >14.0\\
\ion{C}{ii} $\lambda$1334  &>15.9 &>15.5 & >16.1\\   
\ion{C}{ii$^*$} $\lambda$1335 &>14.9 &>14.8 & >15.1\\
\ion{Fe}{ii} $\lambda$1608,  $\lambda$1611, $\lambda$2344, $\lambda$2374, $\lambda$2382, $\lambda$2586, $\lambda$2600 &>14.6 &15.00$\pm$ 0.02 &>15.1\\ 
\ion{Fe}{ii$^*$} $\lambda$2333, $\lambda$2365, $\lambda$2389, $\lambda$2396 &13.09  $\pm$ 0.19 &13.27  $\pm$ 0.14 & 13.49  $\pm$ 0.11\\ 
\ion{Mg}{i} $\lambda$2852 &12.18 $\pm$ 0.19 &12.13 $\pm$ 0.16 & 12.46 $\pm$ 0.13\\ 
\ion{Mg}{ii} $\lambda$2796, $\lambda$2803&>16.0 &>14.3 & >16.0\\   
\ion{Ni}{ii}  $\lambda$1317, $\lambda$1370, $\lambda$1454, $\lambda$1709, $\lambda$1741 &13.39 $\pm$ 0.14 &13.96 $\pm$ 0.06 & 14.06 $\pm$ 0.05\\ 
\ion{O}{i}   $\lambda$1302 &>15.9 &>15.9 & >16.2\\    
\ion{O}{i*}   $\lambda$1304 & &13.97 $\pm$ 0.19 & 13.97 $\pm$ 0.19\\ 
\ion{S}{ii} $\lambda$1250, $\lambda$1253, $\lambda$1259 &15.30$\pm$ 0.10 &15.45$\pm$ 0.06 & 15.68 $\pm$ 0.10\\ 
\ion{Si}{ii} $\lambda$1260, $\lambda$1304, $\lambda$1527, $\lambda$1808 &15.66$\pm$0.02 &15.84$\pm$0.02 & 16.06$\pm$0.01\\   
\ion{Si}{ii$^*$}  $\lambda1264$, $\lambda1265$, $\lambda1309$, $\lambda1533$, $\lambda1817$, $\lambda1818$ &13.90$\pm$0.09 &13.79$\pm$0.05 & 14.15$\pm$0.05\\
\ion{Zn}{ii}  $\lambda$2026 &12.79 $\pm$ 0.03 &13.23 $\pm$ 0.02 & 13.36 $\pm$ 0.02\\ 
\hline
\hline
$b\,{\rm (km~s^{-1})}$ & $20\pm2$ & $34\pm2$ & \\
\hline
\hline
\hline
\hline
\ion{Al}{iii} $\lambda$1854, $\lambda$1862 &13.10 $\pm$ 0.02 &12.66 $\pm$ 0.05 & 13.22 $\pm$ 0.02\\
\ion{C}{iV} $\lambda$1548, $\lambda$1550 &>14.6 &13.31$\pm$ 0.05 & >14.6\\   
\ion{Si}{iV} $\lambda$1393, $\lambda$1402 &>14.4 &12.94$\pm$ 0.07 & >14.5\\  
\hline
\hline
$b\,{\rm (km~s^{-1})}$ & $17\pm2$ & $23\pm2$ & \\
\hline
\hline
    \end{tabular}
    \tablefoot{Column densities of the two strongest components (second and third columns) and the corresponding total column densities of  the low (upper panel), and high ionization (bottom panel) absorption lines of the RRM5 observation.}
       \label{Column_Dens_RRM5}
\end{table*}

\begin{table}[h!]
\caption{Epoch 2 strongest components' column densities}
\scriptsize
\centering
    \begin{tabular}{lcccc}
    \hline
Absorption lines & $\log(N_{I}/{\rm cm}{^{-2}})$  & $\log(N_{II}/{\rm cm}{^{-2}})$ & $\log(N_{TOT}/{\rm cm}{^{-2}})$ \\
\hline
\ion{Al}{ii} $\lambda$1670 &>13.4 &>13.7 & >13.9\\
\ion{C}{ii} $\lambda$1334  &>15.4 &>15.7 & >15.9\\   
\ion{C}{ii$^*$} $\lambda$1335 &>14.6 &>14.7 & >15.0\\
\ion{Fe}{ii} $\lambda$1608,  $\lambda$1611, $\lambda$2344, $\lambda$2374, $\lambda$2382, $\lambda$2586, $\lambda$2600 &>14.5 &15.15$\pm$0.12 &>15.2\\ 
\ion{Fe}{ii$^*$} $\lambda$2333, $\lambda$2365, $\lambda$2389, $\lambda$2396 &13.2  $\pm$ 0.2 &13.2  $\pm$ 0.5 & 13.5  $\pm$ 0.3\\ 
\ion{Mg}{ii} $\lambda$2796, $\lambda$2803&>16.1 &>15.0 & >16.1\\   
\ion{O}{i}   $\lambda$1302 &>16.1 &>15.1 & >16.1\\    
\ion{S}{ii} $\lambda$1250, $\lambda$1253, $\lambda$1259 &15.1$\pm$ 0.2 &15.7$\pm$ 0.2 & 15.8 $\pm$ 0.2\\ 
\ion{Si}{ii} $\lambda$1260, $\lambda$1304, $\lambda$1527, $\lambda$1808 &15.61$\pm$0.12 &15.86$\pm$0.12 &16.05$\pm$0.11\\   
\ion{Si}{ii$^*$}  $\lambda1264$, $\lambda1265$, $\lambda1309$, $\lambda1533$, $\lambda1817$, $\lambda1818$ &13.66$\pm$0.17 &13.45$\pm$0.15 & 13.87$\pm$0.12\\
\ion{Zn}{ii}  $\lambda$2026 &12.54 $\pm$ 0.18 &13.06 $\pm$ 0.10 & 13.17 $\pm$ 0.10\\ 
\hline
\hline
$b\,{\rm (km~s^{-1})}$ & $24\pm2$ & $35\pm2$ & \\
\hline
\hline
\hline
\hline
\ion{Al}{iii} $\lambda$1854, $\lambda$1862 &12.9 $\pm$ 0.2 &12.8 $\pm$ 0.2 & 13.2 $\pm$ 0.2\\
\ion{C}{iV} $\lambda$1548, $\lambda$1550 &>14.4 &>13.4 & >14.5\\   
\ion{Si}{iV} $\lambda$1393, $\lambda$1402 &>14.4 &>12.7 & >14.4\\  
\hline
\hline
$b\,{\rm (km~s^{-1})}$ & $17\pm2$ & $27\pm2$ & \\
\hline
\hline
    \end{tabular}
    \tablefoot{Column densities of the two strongest components (second and third columns) and the corresponding total column densities of  the low (upper panel), and high ionization (bottom panel) absorption lines of the Epoch 2 observation.}
       \label{Column_Dens_Epoch2}
\end{table}

\end{appendices}

\end{document}